\documentclass[useAMS,usenatbib,onecolumn]{mnras}
\usepackage{amsmath}
\usepackage{graphicx,txfonts,fleqn,enumerate,color}
\usepackage{color}
\usepackage{ulem}

\newcommand*  {\diff}    {\mathop{}\!\mathrm{d}}
\renewcommand*{\vec}[1]  {\boldsymbol{#1}}
\renewcommand*{\v}[1]  {\boldsymbol{#1}}

\newcommand*  {\s}[1]    {\mathsf{#1}}
\newcommand*  {\sv}[1]   {\vec{\s{#1}}}
\newcommand*  {\p}       {\partial}
\newcommand*  {\Exp}[1]  {\mathrm{e}^{#1}}

\newcommand*  {\phm}     {\phantom{-}}
\newcommand*  {\op}[1]   {{\hat{#1}}}

\newcommand*  {\Vp}      {V_{\mathrm{pl}}}

\newcommand   {\sub}[2]  {{#1}_{\mathrm{#2}}}
\newcommand*  {\Tp}      {\sub{T}{pl}}
\newcommand*  {\Tpt}      {\sub{\tilde{T}}{pl}}
\newcommand*  {\Tc}      {\sub{T}{ch}}

\newcommand*  {\tdiff}[2]{\frac{\diff{#1}}{\diff{#2}}}
\newcommand*  {\pdiff}[2]{\frac{\p{#1}}{\p{#2}}}

\newcommand*  {\twovector}[2] {{\begin{pmatrix} $1 \\ $2 \end{pmatrix}}}

\renewcommand {\emph}[1]  {\textit{#1}}

\title[Symplectic integrators for planetary systems]
	  {A study of symplectic integrators for planetary system problems: error analysis and comparisons}
\date{Accepted .
      Received ;
      }
\author[ David M.\ Hernandez \& Walter Dehnen]
	   {David M. Hernandez$^1$%
	   \thanks{dmhernan@mit.edu,wd11@le.ac.uk},
	   Walter Dehnen$^2$\footnotemark[1]
	   \\
	   $^1$Department of Physics and Kavli Institute for Astrophysics and Space Research, Massachusetts Institute of Technology,\\
	   	   \phantom{$^1$}77 Massachusetts Ave., Cambridge, Massachusetts 02139, USA \\
	   $^2$Department of Physics \& Astronomy, University of Leicester,
	   	Leicester, LE1 7RH, UK\\
	  }
	
\pubyear{2016}
\begin{document}
\maketitle
\begin{abstract}
{The symplectic Wisdom-Holman map revolutionized long-term integrations of planetary systems. There is freedom in such methods of how to split the Hamiltonian and which coordinate system to employ, and several options have been proposed in the literature.  These choices lead to different integration errors, which we study analytically and numerically.  The Wisdom-Holman method in Jacobi coordinates and the method of Hernandez, H16, compare favorably and avoid problems of some of the other maps, such as incorrect center-of-mass position or truncation errors even in the one-planet case.  We use H16 to compute the evolution of Pluto's orbital elements over 500 million years in a new calculation.}
\end{abstract}

\begin{keywords}
gravitation - methods: analytical - methods: numerical - celestial mechanics - planets and satellites: dynamical evolution and stability
\end{keywords}

\section{Introduction}
\label{sec:int}
Symplectic integrators first became popular in the 1990's \citep{Yoshida1990,chan90} and they revolutionized our ability to study chaotic Hamiltonian dynamical systems for long times.  For long term planetary studies such as investigating chaos in our solar system or stability of newly discovered planetary systems they have become the standard algorithms for solving the orbital ordinary differential equations.  We are concerned here merely with orbital motions; and to first approximation we can treat the system as an $N$-body problem \citep{HeggieHut2003}: $N$ point particles interacting through Newton's gravity.  Explicit symplectic integrators can be interpreted as simply modifying rapidly oscillating terms in the $N$-body Hamiltonian \citep{WH91}. 

Symplectic integration works in a broad, but still limited, number of problems: they are used to solve Hamiltonian problems and their associated linear first order differential equations.  Most of the theory regarding symplectic integrators has been developed for time independent Hamiltonian systems, but one could imagine considering time dependent problems as well.  The strength of symplectic integrators is that they tightly constrain the geometry, not the trajectory of bodies.  If the time step $h$ is small enough, symplectic integrators have an associated quantity $\tilde{H}$, closely related to the Hamiltonian $H$, 
\begin{equation}
\tilde{H} = H + H_{\mathrm{err}}(h) = H + h^n H_n + \mathcal O(h^{n+1}),
\label{eq:cons}
\end{equation}
\citep{HLW06}, where $H_{\mathrm{err}} $ is known as the error Hamiltonian and can be written as a power series in $h$, and $n$ is the order of the integrator.  $\tilde{H}$ is also known as the surrogate Hamiltonian.  {It can be shown there exists a truncation of \eqref{eq:cons} at some power of $h$ that is nearly conserved \citep{HLW06}.}  For planetary system problems, an $N$-body problem described by linear first order ODE's, an $h$ that is  a fraction, say $5\%$, of the shortest orbital period is sufficiently small \citep{W15} in the absence of close encounters.  

Because symplectic solar system integrators for problems with a dominant mass are so commonly used in the literature, here we wish to present, compare, and contrast some major methods.  This paper has a number of new contributions.  We derive and test the error terms for the methods.  We {restore the full Hamiltonian to rederive the} canonical heliocentric map of  \cite{LaskarRobutel1995} and demonstrate its similarity to WHI, the Wisdom-Holman method in inertial Cartesian coordinates, which we have not seen used before.  We show the superior performance of the new H16 and HB15 maps \cite{Hernandez2016,HB15}, although they are more expensive than the other maps in this paper.  We derive and test the {correct} dependence of these methods on the ratio of the typical planetary to solar mass $\epsilon${, which is not typically discussed in the literature}.  We carry out two comparison tests: first we compare the energy error of a planetary system composed of the Sun and outer gas giant planets, and second we add Pluto to the previous problem and compute the error in its inclination for the maps.  The method of Wisdom-Holman in Jacobi coordinates \citep{WH91} and H16 compare favorably.  We use the latter to carry out a new calculation of the orbital elements of Pluto as a function of time over 500 million years, in a repeat of an experiment done in \citep{WH91}.  Some behavior is modified due to different initial conditions.  

The paper is organized as follows: in Section \ref{sec:prelim} we review preliminaries: first we review how to construct symplectic integrators using the operator splitting method and then discuss the $n$-planet problem, a special type of $N$-body problem.  Section \ref{sec:can} derives the necessary coordinate systems needed for the rest of the paper.  In Section \ref{sec:symp} we derive planetary system symplectic maps.  Comparisons between the methods and our calculation of Pluto's orbital elements over time are carried out in Section \ref{sec:comp}.  We conclude in Section \ref{sec:conc}.  

The maps in this paper do not handle close encounters between planets {\citep[except for HB15, see][]{Hernandez2016}}.  Two popular methods for incorporating close encounters are the integrators of \cite{Chambers1999}, \textsc{MERCURY} and \cite{DuncanLevisonLee1998}, \textsc{SyMBA}: both are based on the Wisdom-Holman method in Democratic Heliocentric coordinates (Section \ref{sec:WHD}).  For all the tests in this paper, we use the Kepler solver of \cite{WisdomHernandez2015}.
\section{Preliminaries}
\label{sec:prelim}
\subsection{Symplectic Integrators from Operator Splitting}
\label{sec:op}
Consider a Hamiltonian $H$ with canonical coordinates $\v{w} = (\v{x},\v{p})$, where $\v{p}$ are the momenta conjugate to $\v{x}$.  We define the action of an operator $\hat{H}$ on an arbitrary function $\v{f}(\v{w})$ as $\hat{{H}} \v{f}(\v{w}) = \{ \v{f}(\v{w}), {{H}(\v{w})}\}$, where $\{,\}$ denotes a Poisson bracket.  Then, Hamilton's equations say
\begin{equation}
\v{\dot{w}} = \hat{H} \v{w}.
\end{equation}
If $H$ has no explicit time-dependence, the formal solution to this differential equation is that the new phase space coordinates $\v{w}^\prime$ after time $h$ are 
\begin{equation}
\label{eq:exact}
\v{w}^\prime = \Exp{h \hat{H}} \v{w}.
\end{equation}    
If $C(\v{w}) = A(\v{w}) + B(\v{w})$, then $\hat{C} = \hat{A} + \hat{B}$.  A Hamiltonian is integrable if the initial value problem corresponding to the Hamiltonian can be solved by integrals over known functions.  For example, the Kepler or simple harmonic oscillator Hamiltonian is integrable.  \footnote{The corresponding ODE's for these problems are in fact the same through a coordinate and time transformation \citep{Aarseth2003}.}  It is convenient to split non-integrable Hamiltonians into a sum of integrable Hamiltonians when possible.  A judicious choice of coordinates can help solve the split Hamiltonians or even determine if they are integrable.  For example, for the $N$-body Hamiltonian in Cartesian coordinates, $H = T(\v{p}) + V(\v{x})$, and $T(\v{p})$ and $V(\v{x})$ are both integrable Hamiltonians, even though $H$ is not.  Assuming $H$ is non-integrable, if $H = A(\v{w}) + B(\v{w})$, with $A$ and $B$ integrable, we can write \citep[][and references therein]{HLW06}
\begin{equation}
\label{eq:BCH}
\exp{\left(h\hat{H} \right)} \approx \exp{\left(h\hat{\tilde{H}}\right)} \equiv \exp{\left(h\op{A} \right)}\exp{\left(h\op{B} \right)} = \exp\left(h(\hat{A}+\hat{B}) + \tfrac{h^2}{2}[\hat{A},\hat{B}] + \tfrac{h^3}{12}\big([\hat{A},[\hat{A},\hat{B}]] + [\hat{B},[\hat{B},\hat{A}]]\big) + 
	\;\dots \right),
\end{equation}
where $[\hat{A},\hat{B}]\equiv \hat{A} \hat{B}-\hat{B} \hat{A}$ is the commutator of $\hat{A}$ and $\hat{B}$ . Equation \eqref{eq:BCH} uses the so called Baker-Campbell-Hausdorff (BCH) formula and is a first order mapping, the simplest we can write, that can be used to approximately solve $H$ explicitly.  The integrator dynamics are described not by $H$, but by $\tilde{H}$, which has been shown to be approximately conserved over exponentially long times \citep{HLW06}.  For the $N$-body problem we could let $\hat{A} = \hat{T}$ and $\hat{B} = \hat{V}$  Specifically, the new phase space coordinates after small time $h$ are found from 
\begin{equation}
\label{eq:weq}
\v{w}^\prime = \Exp{h \hat{T}}\,\Exp{h \hat{V}} \v{w},
\end{equation}
or
\begin{equation}
\label{eq:simpmap}
\v{p}^\prime = \v{p} - h \frac{\partial V}{\partial \v{x}}(\v{x}) \qquad \text{and} \qquad \v{x}^\prime = \v{x} + h \frac{\partial T}{\partial \v{p}}(\v{p}^\prime).
\end{equation}
Eqs. \eqref{eq:simpmap} solve the $N$-body problem approximately.  The first equation of \eqref{eq:simpmap}, which modifies momentum while leaving the position invariant, is a `kick,' while the second is a `drift.'  Equations \eqref{eq:weq} or \eqref{eq:simpmap} are known as a symplectic Euler method.  

{Now we} define $\sv{J} = \partial \v{w}^\prime/ \partial \v{w}$ and let $\sv{\Omega}$ be a constant matrix whose form depends on the ordering of the phase space coordinates in $\v{w}$.  A phase space map is symplectic if 
\begin{equation}
\label{eq:sympcond}
\sv{\Omega} = \v{J}^\dag \sv{\Omega} \sv{J}.
\end{equation}
For the ordering $\v{w} = (\v{x},\v{p})$,  
\begin{equation}
\sv{\Omega} = 
	\begin{bmatrix}
	\sv{0}_m & {\sv{I}}_m \\
	-{\sv{I}}_m & \sv{0}_m \\
	\end{bmatrix}.
\end{equation} 
$2 m$ is the dimension of $\v{w}$, and $\sv{0}_m$ and ${\sv{I}}_m$ are the square zero and identity matrices, respectively, of size $m \times m$.  Eq. \eqref{eq:sympcond} gives $2m^2 + m$ independent constraints, demonstrating symplecticity is a stringent requirement on the phase space geometry.  {These constraints do not reduce the dimensionality of phase space, they are constraints on differential forms of phase space.}  

The map \eqref{eq:simpmap} satisfies \eqref{eq:sympcond}.   The problem is that \eqref{eq:simpmap} is usually not accurate enough for various reasons: first, this approximation is often too low an order and is not time-symmetric.  Next, other Hamiltonian splittings besides $H = T(\v{p}) + V(\v{x})$ are potentially superior {as far as error analysis of the map is concerned}.  Finally, we have not specified our choice of canonical coordinates; some choices are more practical.  Maps such as \eqref{eq:BCH} and the others we are interested in this paper are \textit{one-step methods}, meaning each phase space map is an initial value problem and does not have memory of past solutions.  They are arguably the simplest kind to implement.  Using the Jacobi identity
\begin{equation}
	\{\{A,B\},C\} + \{\{B,C\},A\} + \{\{C,A\},B\} = 0,
\end{equation}
combined with \eqref{eq:BCH}, we find the surrogate Hamiltonian is
\begin{equation}\label{eq:htild}
	\tilde{H} =
	A + B + 
	\frac{h}{2} \{B,A\} +
	\frac{h^2}{12} \{\{B,A\},A\} +
	\frac{h^2}{12} \{\{A,B\},B\} +
	\mathcal{O}(h^3):
\end{equation}
we see the method is first order (Section \ref{sec:int}).  This function is approximately conserved \citep{HLW06} and its operator is the time evolution operator, as seen from \eqref{eq:BCH}. The Jacobi identity implies that if two functions are in involution, the commutator of their operators is zero; in equations, if $\{A,B \} = 0$, then $[\hat{A},\hat{B}] = 0$ and $ \exp\left(\hat{A}+\hat{B} \right) = \Exp{h \hat{A}}\,\Exp{h \hat{B}}$: the order of application of the $\hat{A}$ and $\hat{B}$ maps does not matter.  While $T$ and $V$ above are not in involution, other functions from other Hamiltonian splittings are, as we will see in later sections.

The program we follow in this paper is to consider some Hamiltonian splitting such as $H = A(\v{w}) + B(\v{w})$, with associated operators $\hat{A}$ and $\hat{B}$, and use these operators to construct a map such as eq. \eqref{eq:BCH}.  The error Hamiltonian $\tilde{H}$ is computed using the procedure in this section or an analogue \citep{HLW06} for more complicated maps such as 
\begin{equation}
\label{eq:symm}
\Exp{h \hat{\tilde{H}}} = \Exp{\frac{h}{2} \hat{B}} \Exp{h \hat{A}}\Exp{\frac{h}{2} \hat{B}},
\end{equation}
which is second order.  As long as $h$ is small enough, a higher order integrator indicates the effective Hamiltonian is more accurate and consequently so are the phase space variables.  The integrators are usually only second order; higher order methods do not make sense for long term studies because these integrators are optimized for high precision, but chaos prevents conclusions based on exact trajectories \citep{Hernandez2016}.  {Some authors \citep[e.g.][]{Fetal13} have proposed high order symplectic integrators, but their usefulness is unclear.  For studying exact trajectories over short times superior non-symplectic options such as that of \cite{RS15} already exist.}

The adjoint of a one step method $\phi_h$ is simply $\phi_{h}^\dagger = \phi_{-h}^{-1}$.  If $\phi_h = \phi_h^\dagger$, the integrator is \textit{time-reversible}.  Eq. \eqref{eq:symm} is time-reversible while \eqref{eq:BCH} is not when $\{A,B\} \ne 0$.  We can write a symplectic integrator from operator splitting as 
\begin{equation}
\phi_h = \exp \left(\gamma_1 h + \gamma_2 h^2 + \hdots \right),
\end{equation}
with $\gamma_i$ some function of operators.  Using the BCH equation, the time reversibility condition, which also reads $\phi_h \phi_{-h} = \sv{I}_{2 m}$, would require $\gamma_i = 0$ for $i$ even.  Time reversibility is a desirable property for integrators \citep{HLW06,HutMakinoMcMillan1995}; it is also desirable for symplectic integrators to be a low, say second or fourth, order (Section \ref{sec:int}).  Thus, we will frequently encounter the form \eqref{eq:symm} in this paper.    

\subsection{\boldmath{The $n$-planet problem}}
Consider a dynamical system consisting of a single massive particle, the `Sun', surrounded by $n$ particles of considerably lower masses, the `planets'. Let $m_i$ be the particle masses with $i=0$ denoting the Sun, $\vec{x}_i$ their inertial Cartesian coordinates, and $\vec{p}_i\equiv m_i\dot{\vec{x}}_i$ their conjugate momenta. Then the Hamiltonian of the system is
\begin{equation}
	H = T + V = \sum_{i} \frac{\vec{p}_i^2}{2m_i} - \sum_{i<j}\frac{Gm_im_j}{r_{i j}},
\end{equation}
where $r_{i j} = |\vec{x}_i-\vec{x}_{\!j}|$.  This Hamiltonian has $6n+5$ functions of phase space that are constant along the trajectory, but, in practice, only 7 actually reduce the dimensionality of the phase space and are thus called isolating integrals.  They are $H$ , $\vec{p}_{\mathrm{tot}}=\sum_i \vec{p}_i$ , and $\vec{L} = \sum_i {\vec{x}_i} \times \vec{p}_i${, and they} have vanishing Poisson bracket with the Hamiltonian.  The vector $\vec{R} =  \sum_i m_i {\vec{x}}_i -t \sum_i \vec{p}_i$ has $\dot{\v{R}} = 0$, is obtained by integrating $\vec{p}_{\mathrm{tot}}=\sum_i \vec{p}_i$ with $3n + 3$ initial conditions.  
For more general, not necessarily Cartesian, canonical coordinates $\vec{q},\vec{p}, $ the $n = 1$ Hamiltonian has 6 degrees of freedom and is written
\begin{equation}
\label{eq:neq1}
\label{eq:n1Hamilt}
	H =  \frac{{P}_0^2}{2 M} + \frac{{P}_1^2}{2 \mu} - \frac{G M \mu}{Q_1},
\end{equation}
where $\vec{P}_1 = \pm (\vec{p}_1 m_0 - \vec{p}_0 m_1)/M $,  $\vec{Q}_1 = \pm (\vec{q}_0-\vec{q}_1)$, $\vec{P}_0 = \vec{p}_0 +\vec{p}_1$, $\mu$ and $M$ are the reduced and total mass, respectively, and the magnitude of a vector $\v{A}$ is written $A$.  {If $P_0 = 0$, $\v{P}_1 = \pm \v{p}_1 = \mp \v{p}_0$.}  Only this $n = 1$ problem is integrable.  But actually there also exists a conserved Laplace-Runge-Lenz vector,
\begin{equation}
\v{A} = \v{P_1} \times \left(\v{Q}_1 \times \v{P}_1 \right) - G \mu^2 M \frac{\v{Q}_1}{Q_1}, 
\end{equation}
which implies $\{\v{A},H\} = 0$ and $A/G \mu^2 M = e$ for $e \ne 0$, with $e$ the eccentricity.  $0 \le e < 1$ for elliptical (bound) orbits, $e > 1$ for hyperbolic (unbound) orbits, and $e = 1$ for parabolic (zero energy) orbits.  \eqref{eq:n1Hamilt} is called \textit{superintegrable} because there are {8 integrals}.  Although we are concerned with $n > 1$ in this paper, it is useful to think of planetary orbits as nearly obeying $n = 1$ problems with the Sun.  We generally assume $e < 1$ in these planetary orbits, which is the more interesting case because we must calculate more than one orbit.  But the maps we describe in this paper work for any $e$. 

For approximating the solution for $n > 1$, it is often beneficial to consider other coordinate systems apart from inertial Cartesian coordinates, which we discuss in Section \ref{sec:can}.  Unfortunately, some existing methods for solving $n > 1$ problems do not solve the $n = 1$ problem exactly; we discuss how to fix this problem.  

\section{\boldmath Canonical Coordinate Systems for the $n$-planet problem}
\label{sec:can}
\subsection{Heliocentric Coordinates}
To define heliocentric coordinates, let
\begin{subequations}
	\label{eqs:Q,P}
\begin{equation}
	\label{eq:Qi}
	\vec{Q}_{i\neq0} = \vec{x}_i-\vec{x}_0.
\end{equation}
However, it is not immediately obvious what to take for $\vec{Q}_0$. Let us therefore allow a general linear form
\begin{equation}
	\label{eq:Q0}
	\vec{Q}_0 = \sum_i a_i\vec{x}_i,
\end{equation}
where the weights $a_i$ satisfy 	$\sum_ia_i=1$ without loss of generality. 
If the associated canonical momentum vector is $\v{P}$, the fundamental Poisson brackets satisfy $\{\vec{Q}_i,\vec{Q}_{\!j}\}=\sv{0}_3=\{\vec{P}_i,\vec{P}_{\!j}\}$ and $\{\vec{Q}_i,\vec{P}_{\!j}\}= \delta_{i\!j} \sv{I}_3$ as required (and sufficient) for a canonical transformation.  Explicitly, 
\begin{equation} \label{eq:P:of:p}
	\vec{P}_0 = \sum_j \vec{p}_{\!j}
	\qquad\text{and}\qquad
	\vec{P}_{i\neq0} = \vec{p}_i - a_i \sum_j \vec{p}_{\!j}.
\end{equation}
These relations have inversion
\begin{eqnarray}
	\vec{x}_0 &=& \vec{Q}_0 - \sum_{j\neq0} a_{\!j}\,\vec{Q}_{\!j}
	\qquad\text{and}\qquad
	\vec{x}_{i\neq0} = \vec{Q}_0 + \vec{Q}_i - \sum_{j\neq0} a_{\!j}\,\vec{Q}_{\!j},
	\\
	\label{eq:p:of:P}
	\vec{p}_0 &=& a_0 \vec{P}_0 - \sum_{j\neq0}\vec{P}_{\!j}
	\qquad\text{and}\qquad
	\vec{p}_{i\neq0} = a_i\vec{P}_0 + \vec{P}_i.
\end{eqnarray}
\end{subequations}
In particular, $\vec{P}_0$ is always equal to the total momentum
\begin{equation}
	\sub{\vec{p}}{tot} \equiv \sum_{i}\vec{p}_i,
\end{equation}
regardless of the choice for $a_i$. Thus, the weights $a_i$ only affect $\vec{Q}_0$ and $\vec{P}_{i\neq0}$.  The total angular momentum has the same functional dependence on the new as on the old coordinates,
\begin{equation}
\label{eq:ang}
	\sub{\vec{L}}{} \equiv \sum_{i} \vec{x}_{i}\times\vec{p}_{i}
	= \sum_{i} \vec{Q}_i \times \vec{P}_i.
\end{equation}
In fact, we can check for any maps linear in the coordinates and momenta, $\vec{Q}_i = \sum_j {A}_{i j} \vec{x}_j$ and $\vec{P}_i = \sum_j B_{i j} \vec{p}_j$, the form of the angular momentum is unchanged, and the canonical transformations require ${\sv{B}}^\dagger = {\sv{A}}^{-1}$.  The kinetic and potential energies as function of the new coordinates and momenta are
\begin{subequations}
	\label{eq:T,V:a}
\begin{eqnarray}
	T &=& \left(\sum_{i} \frac{a_i^2}{2m_i}\right) \vec{P}_0^2
	  + \sum_{i\neq0} \frac{\vec{P}_i^2}{2m_i}
	  + \frac{1}{2m_0} \left(\sum_{i\neq0}\vec{P}_i\right)^2
	\label{eq:T:a}  +\,
	     \vec{P}_0 \cdot
	   	 \sum_{i\neq 0}\left(\frac{a_i}{m_i}-\frac{a_0}{m_0}\right) \vec{P}_i,
	\\[0.5ex] \label{eq:V:a}
	V &=& - \sum_{i\neq0} \frac{Gm_im_0}{|\vec{Q}_i|}
	    - \sum_{0<i<j} \frac{Gm_im_{\!j}}{|\vec{Q}_{i}-\vec{Q}_{\!j}|}.
\end{eqnarray}
\end{subequations}

\subsubsection{Democratic Heliocentric Coordinates}
The best choice for the weights is such that the last term in equation~(\ref{eq:T:a}) vanishes for any value of the $\vec{P}_i$. Together with the normalisation constraint $\sum_ia_i=1$ this leads uniquely to
\begin{equation}
	a_i = \frac{m_i}{M}
	\quad\text{with}\quad
	M \equiv {\sum_im_i},
\end{equation}
and the kinetic energy becomes
\begin{eqnarray} \label{eq:DHC:T}
	\label{eq:T:d:a}
	T &=& \frac{\vec{P}_0^2}{2M}
	  + \sum_{i\neq0} \frac{\vec{P}_i^2}{2m_i}
	  + \frac{1}{2m_0} \left(\sum_{i\neq0}\vec{P}_i\right)^2
\end{eqnarray}
The relations~(\ref{eqs:Q,P}) become
\begin{subequations}
	\label{eqs:DHC:Q,P}
\begin{align}
	\vec{Q}_0 &= \sum_j\frac{m_j}{M}\vec{x}_{\!j},
	&
	\vec{Q}_{i\neq0} &= \vec{x}_i-\vec{x}_0,
	\\
	\vec{P}_0 &= \sum_j\vec{p}_{\!j},
	&
	\vec{P}_{i\neq0} &= \vec{p}_i - \frac{m_i}{M}\vec{P}_0 = 
		\vec{p}_i - \frac{m_i}{M}\sum_j\vec{p}_{\!j},
	\\
	\vec{x}_0 &= \vec{Q}_0-\sum_{j\neq0}\frac{m_j}{M}\vec{Q}_{\!j},
	&
	\vec{x}_{i\neq0} &= \vec{x}_0+\vec{Q}_i = 
		\vec{Q}_0+\vec{Q}_i-\sum_{j\neq0}\frac{m_j}{M}\vec{Q}_{\!j},
	\\
	\vec{p}_0 &= \frac{m_0}{M}\vec{P}_0 - \sum_j\vec{P}_{\!j},
	&
	\vec{p}_{i\neq0} &= \frac{m_i}{M}\vec{P}_0 + \vec{P}_i.
\end{align}
\end{subequations}
Thus, $\vec{Q}_0$ is the centre of mass and $\vec{P}_{i\neq0}$ the barycentric momenta of the planets. These coordinates have been introduced by \cite*{DuncanLevisonLee1998}, who dubbed them `democratic heliocentric coordinates'. \cite{W06} refers to these coordinates as `canonical heliocentric coordinates,' but we give priority to the naming of \cite{DuncanLevisonLee1998}.  For these coordinates $M\dot{\vec{Q}}_0=\vec{P}_0$ and $\ddot{\vec{Q}}_0=0$.

It is often convenient to work with velocities rather than momenta. The barycentric velocities
\begin{equation}
	\tilde{\vec{v}}_i\equiv\vec{v}_i-\sub{\vec{v}}{\!cm},
\end{equation}
where
\begin{equation}
	\sub{\vec{v}}{\!cm}\equiv \vec{P}_0/M
\end{equation}
is the centre-of-mass velocity.  The barycentric velocities $\tilde{\v{v}}_i$ are related to the momenta $\vec{P}_i$ via
\begin{equation}
	\tilde{\vec{v}}_0 = - \frac{1}{m_0}\sum_{j\neq0}\vec{P}_{\!j}
	\quad\text{and}\quad
	\tilde{\vec{v}}_{i\neq0} = \frac{\vec{P}_i}{m_i}.
\end{equation}
The kinetic energy \eqref{eq:T:d:a} can then be rewritten
\begin{equation}
T =\frac{1}{2}M\sub{\vec{v}}{\!cm}^2 
	  + \sum_{i}\frac{1}{2}m_i\tilde{\vec{v}}_i^2.
\end{equation}

\subsubsection{Canonical Heliocentric Coordinates}
\label{sec:chcoord}
Another obvious choice for the weights is $a_i=\delta_{i0}$ \citep{Poincare1896, LaskarRobutel1995}, such that $\vec{Q}_0=\vec{q}_0$ remains the Solar position and $\vec{P}_{i\neq0}=\vec{p}_i$ the planetary inertial momenta. \citeauthor{LaskarRobutel1995} dubbed these `canonical heliocentric coordinates'.  Following \citeauthor{LaskarRobutel1995} we call these `canonical heliocentric coordinates' (note that \cite{W06} used this name for the democratic heliocentric coordinates).  The kinetic energy becomes 
\begin{eqnarray}
	\label{eq:T:c:a}
	T&=& \frac{\vec{P}_0^2}{2m_0}
	  + \sum_{i\neq0} \frac{\vec{P}_i^2}{2m_i}
	  + \frac{1}{2m_0} \left(\sum_{i\neq0}\vec{P}_i\right)^2
	  - \frac{\vec{P}_0}{m_0} \cdot \sum_{i\neq0} \vec{P}_i.
\end{eqnarray}
For these coordinates $\dot{\vec{Q}}_0$ is not related to $\vec{P}_0$ and
\begin{equation}
	\label{eq:o:m:Q:c:0}
	\ddot{\vec{Q}}_{0} = \ddot{\vec{x}}_{0} =
	  \sum_{i\neq0} \frac{Gm_i}{|\vec{Q}_i|^3}\vec{Q}_i,
\end{equation}
is the inertial acceleration of the Sun. 
Eq. \eqref{eq:T:c:a} differs from (\ref{eq:T:d:a}) in two ways. First, the term quadratic in $\vec{P}_0$ in eq. \eqref{eq:T:c:a} has a different normalisation. However, in the common situation of using a barycentric frame (where $\vec{P}_0=0$) this term has no effect and can be ignored.

The second difference is the presence of a term linear in $\vec{P}_0$ in equation~(\ref{eq:T:c:a}), which was eliminated by design for the democratic heliocentric coordinates. Even though this term vanishes in the barycentric frame, it must be retained when using $T$ as (part of) a Hamiltonian, for otherwise the canonical equation of motion incorrectly obtains $\dot{\vec{Q}}_0=0$. With this term, the canonical equation of motion gives $\dot{\vec{Q}}_0 = \dot{\vec{x}}_0$ as required.  Similarly, for the centre-of-mass position
\begin{equation}
	\sub{\vec{x}}{cm} = \sum_i \frac{m_i}{M} \vec{x}_i
	= \left[\frac{m_0}{M}\vec{Q}_0 +
		\sum_{i\neq0} \frac{m_i}{M} (\vec{Q}_0+\vec{Q}_i)\right]
	= \vec{Q}_0 + \sum_{i\neq0} \frac{m_i}{M}\vec{Q}_i
\end{equation}
we get the correct $\sub{\dot{\vec{x}}}{cm} = {\vec{P}_0}/{M}$ 
while without the last term in equation~(\ref{eq:T:c:a}), $\sub{\dot{\vec{x}}}{cm} =({2\vec{P}_0-\vec{p}_0})/{m_0}$.  \cite{LaskarRobutel1995} used the form~(\ref{eq:T:d:a}) instead of (\ref{eq:T:c:a}) with the canonical heliocentric coordinates, i.e.\ neglected the last term in equation~(\ref{eq:T:c:a}).  {This omission affects the Solar motion (and hence all barycentric trajectories) and the heliocentric orbits if we do not work in a barycentric frame.}  Even {when} using the {full} form ~(\ref{eq:T:c:a}) of the kinetic energy, the center of mass position will be incorrect in our symplectic integrators, as we will see.  {Fortunately, in all these situations, the} error {made} can be simply corrected by adjusting $\v{Q}_0$ {using the conservation laws (e.g. \cite{Fetal13})}.
\subsection{Jacobi Coordinates}
\label{sec:jac}
Canonical Jacobi coordinates are well-suited to studying the $n$-planet problem.  They are constructed by first defining a system of relative and center of mass positions.  The canonical momenta then follow by requiring a symplectic coordinate system.  We study the appropriate coordinate system for problems with a dominant mass with index 0.  The planets are ordered according to the distance from the Sun.  Other coordinate systems can be derived for problems with more than one dominant mass assuming the system is stable; these coordinates are called Hierarchical Jacobi Coordinates by \cite{SW01}, who also introduce Jacobi coordinates.  

$\vec{u}_i$ and $\vec{s}_i$ are the canonical position and momentum of planet $i$, respectively, in Jacobi coordinates.  $\vec{g}_i$ and $\vec{G}_i$ are the center of mass and total momentum, respectively, of the system of particles with indices $\le i$.  $M_i = \sum_{j = 0}^i m_i$.  Let $\vec{g}_0 = \vec{x}_0$ and $\vec{G}_0 = \vec{p}_0$.  Then  
\begin{subequations}
\label{eq:tojac}
\begin{align}
\vec{u}_{i \ne 0} &= \vec{x}_i - \vec{g}_{i - 1},
&
\vec{s}_{i \ne 0} &= \frac{M_{i-1}}{M_i} \vec{p}_i - \frac{m_i}{M_i} \vec{G}_{i - 1},
\\
\vec{g}_{i \ne 0} &= \frac{1}{M_i} \left(m_i \vec{x}_i + M_{i - 1} \vec{g}_{i - 1}\right),
&
\vec{G}_{i \ne 0} &= \vec{p}_i + \vec{G}_{i - 1},
\\
\vec{u}_0 &= \vec{g}_{n},
&
\vec{s}_0 &= \vec{G}_{n}.
\end{align}
\label{eq:jac}
\end{subequations}
$\vec{u}_{i \ne 0}$ is the relative position of planet $i$ with respect to the center of mass of bodies with a smaller index and the meaning of $\v{s}_{i \ne 0}$ will be clear below.  $\v{u}_0$ locates the center of mass and $\v{s}_0$ is the total momentum.  For the inverse transformations let $\v{g}_n = \v{u}_0$ and $\v{G}_n = \v{s}_0$.  Then
\begin{subequations}
\label{eq:fromjac}
\begin{align}
\v{x}_{i \ne 0} &= \frac{M_{i - 1}}{M_i} \v{u}_i + \v{g}_{i},
&
\v{p}_{i \ne 0} &= \v{s}_i + \frac{m_i}{M_i} \v{G}_{i},
\\
\v{g}_{i-1, i\ne 0} &= -\frac{m_i}{M_i} \v{u}_i + \v{g}_i,
&
\v{G}_{i-1, i\ne 0} &= \v{s}_i + \frac{M_{i-1}}{M_i} \v{G}_i,
\\
\v{x}_0 &= \v{g}_{0},
&
\v{p}_0 &= \v{G}_{0}.
\end{align}
\label{eq:jacinv}
\end{subequations}
The canonical maps defined by eq'ts. \eqref{eq:jac} and \eqref{eq:jacinv} are linear point transformations.  The following identities are easy to verify:
\begin{equation}
\begin{aligned}
\frac{\v{G}_{i - 1}^2}{2 M_{i - 1}} + \frac{\v{p}_i^2}{2 m_{i}} &= \frac{\v{G}_{i}^2}{2 M_{i}} + \frac{\v{s}_{i}^2}{2 m_i^\prime}, \qquad\text{and}
\\
\v{g}_{i - 1} \times \v{G}_{i - 1} + \v{x}_i \times \v{p}_i &= \v{g}_i \times \v{G}_i + \v{u}_i \times \v{s}_i,
\end{aligned}
\end{equation}  
where $m_{i \ne 0}^\prime = m_i {M_{i - 1}}/{M_i}$ and $m_0 ^\prime = M$. 
We can use them to show the the kinetic energy is diagonal in Jacobi coordinates,
\begin{equation}
T = \frac{\v{s}_0^2}{2 m_0^\prime} + \sum_{i = 1}^{n} \frac{\v{s}_i^2}{2 m_i^\prime},
\label{eq:jacT}
\end{equation}
and that the form of the angular momentum is invariant in Jacobi coordinates:
\begin{equation}
L = \sum \v{x}_i \times \v{p}_i = \sum \v{u}_i \times \v{s}_i,
\end{equation}
though we already know this from the comment under \eqref{eq:ang}. The first term of $T$ drops out when working in a barycentric frame.  The kinetic energy splits into the center-of-mass term plus $n$ (rather than $n+1$) terms, each associated with a planetary orbit, without a remaining term associated solely with the Sun.   

We can use eqs. \eqref{eq:tojac} and \eqref{eq:fromjac} to write more compact maps to and from Jacobi coordinates that bypass the use of $\v{g}_i$ and $\v{G}_i$.  From inertial to Jacobi coordinates,
\begin{subequations}
\label{eq:tojac2}
\begin{align}
\v{u}_{i \ne 0} &= \v{x}_i - \frac{1}{M_{i -1}} \sum_{j = 0}^{i-1} m_j \v{x}_j,
&
\v{s}_{i \ne 0} &= \frac{M_{i-1}}{M_i} \v{p}_i - \frac{m_i}{M_i} \sum_{j = 0}^{i-1} \v{p}_j,
\\
\v{u}_0 &= \frac{1}{M_n} \sum_{j = 0}^n m_j \v{x}_j,
&
\v{s}_0 &= \sum_{j = 0}^n \v{p}_j.
\end{align}
\end{subequations}
(As expected, $\v{s}_{i \ne 0}/m_i^\prime = \dot{\v{u}}_i$).  The inverse transformations are 
\begin{subequations}
\begin{align}
\v{x}_{0 < i < n} &= \v{u}_0 + \frac{M_{i-1}}{M_i} \v{u}_i - \sum_{j = i+1}^n \frac{m_j}{M_j} \v{u}_j,
&
\v{p}_{0< i < n} &= \frac{m_i}{M_n} \v{s}_0 + \v{s}_i - \sum_{j = i+1}^n \frac{m_i}{M_{j - 1}} \v{s}_j,
\\
\v{x}_0 &= \v{u}_0 - \sum_{j = 1}^n \frac{m_j}{M_j} \v{u}_j,
&
\v{p}_0 &= \frac{m_0}{M_n} \v{s}_0 - \sum_{j = 1}^n \frac{m_0}{M_{j - 1}} \v{s}_j.
\\
\v{x}_n &= \v{u}_0 + \frac{M_{n-1}}{M_{n}} \v{u}_n,
&
\v{p}_n &= \frac{m_{n}}{M_n} \v{s}_0 + \v{s}_n.
\end{align}
\end{subequations}
We can use eq'ts. \eqref{eq:tojac2} to check that indeed $\{ \v{u}_i, \v{s}_j \} = \delta_{i j} \sv{I}_3 \forall~ i,j$.  {Unlike the situation for the other coordinates in previous sections, Jacobi coordinates are written assuming no orbit crossings occur.} 
\section{Symplectic maps for planetary systems}
\label{sec:symp}
In planetary systems the motion of each planet is dominated by the attraction of the Sun. It is therefore expedient to integrate this motion with a Kepler solver, but use an ordinary kick-and-drift approach for the inter-planetary interactions. In order to construct a symplectic map with this property via the operator-splitting method, we must re-write the Hamiltonian as $H=A+B$ such that $A$ is sum of $n$ decoupled Kepler problems \citep{WH91}. The only difficulty is that, even if neglecting the inter-planetary forces, the planetary orbits are coupled by their combined tugging at the Sun. In terms of the Hamiltonian, this corresponds to the problem of how to split the Solar kinetic energy.

\subsection{The Wisdom-Holman map in democratic heliocentric coordinates: WHD}
\label{sec:WHD}
\citeauthor{WH91} originally proposed a split based on Jacobi coordinates but a widely used choice which is independent of the ordering of the planets and allows easy addition or removal of planets are democratic heliocentric coordinates \citep*{DuncanLevisonLee1998}.  Here we review this map and derive and test its second order error terms.  We also review its dependence on $\epsilon$, the ratio of planetary to solar mass. 

The kinetic energy is split into
\begin{equation} \label{eq:DHC:T}
	T_0 = \frac{\v{P}_0^2}{2 M} + 
	\frac{1}{2m_0} \left(\sum_{i\neq0}\vec{P}_i\right)^2
	\qquad\text{and}\qquad
	T_1 = \sum_{i\neq0} \frac{\vec{P}_i^2}{2m_i}
\end{equation}
and the potential energy into
\begin{equation} \label{eq:DHC:V}
	V_\odot = -\sum_{i\neq0}\frac{Gm_0m_i}{Q_i}
	\qquad\text{and}\qquad
	\Vp = - \sum_{0<i<j} \frac{Gm_im_{\!j}}{Q_{i\!j}},
\end{equation}
where $\vec{Q}_{i\!j}\equiv\vec{Q}_{i}-\vec{Q}_{\!j}$.   The resulting Wisdom-Holman split of the Hamiltonian is $H=\sub{A}{d}+\sub{B}{d}$, with associated operators $\sub{\hat{A}}{d}$ and $\sub{\hat{B}}{d}$, is
\begin{subequations} \label{eqs:WHD:A+B}
\begin{eqnarray}
	\label{eq:WHD:A}
	\sub{A}{d} &=& T_1 + V_\odot =  
		\sum_{i\neq0}\frac{\vec{P}_i^2}{2m_i} - \frac{Gm_0m_i}{Q_i},
	\\ \label{eq:WHD:BB}
	\sub{B}{d} &=& T_0 + \Vp =
	\frac{\vec{P}_0^2}{2 M} +
		\frac{1}{2m_0}\left(\sum_{i\neq0}\vec{P}_i\right)^2
		- \sum_{0<i<j}\frac{Gm_im_{\!j}}{Q_{i\!j}}.
\end{eqnarray}
\end{subequations}
The WHD maps are then
\begin{equation} 
	\Exp{h \hat{\tilde{H}}} = \Exp{\frac{h}{2} \sub{\hat{B}}{d}} \Exp{h \sub{\hat{A}}{d}}\Exp{\frac{h}{2} \sub{\hat{B}}{d}}
	\qquad\text{or}\qquad
	\Exp{h \hat{\tilde{H}}} = \Exp{\frac{h}{2} \sub{\hat{A}}{d}} \Exp{h \sub{\hat{B}}{d}}\Exp{\frac{h}{2} \sub{\hat{A}}{d}}.
\end{equation}
From \eqref{eq:BCH}, $\Exp{{h} \sub{\hat{B}}{d}} = \Exp{{h} \hat{T}_0} \Exp{{h} \sub{\hat{V}}{pl}}$.  The map $\Exp{{h} \sub{\hat{B}}{d}}$ consists of a drift of each planet by the Solar barycentric reflex motion and a kick by the interplanetary forces. The order of these sub-maps is irrelevant, since the corresponding terms of $\sub{B}{d}$ are in involution. $\sub{A}{d}$ generates Keplerian orbits with gravitating masses $m_0$ such that the total momentum is conserved, but the $n=1$ problem is not solved exactly ($\sub{B}{d} \ne 0$ and $\sub{A}{d} \ne H $). The Kepler problems arising from $\sub{A}{d}$ are mutually independent and can be synchronously computed.    The Poisson bracket of $\v{L}$ or $\v{P}_0$ with $T_1$, $V_\odot$, $T_0$, and $\sub{V}{pl}$ is zero so this map and others based on these functions conserve angular and linear momentum.

\subsubsection{The $\mathcal{O}(h^2)$ error terms}
Computation of $\tilde{H}$ requires calculating Poisson brackets \citep{HLW06} like those in \eqref{eq:htild}.  All possible $\mathcal{O}(h^2)$ terms which may occur in the surrogate Hamiltonian $\tilde{H}$ of a second order map composed of operators $\Exp{h \sub{\op{T}}{1}}$, $\Exp{h \op{V}_\odot}$, $\Exp{h \sub{\op{V}}{pl}}$, and $\Exp{h \sub{\op{T}}{0}}$  are
\begin{subequations}
	\label{eqs:Herr:WHD:split}
\begin{align}
	\{\{V_\odot,T_0\},T_0\}
	&= \sum_{i\neq0}\frac{Gm_i}{m_0Q_i^5}
	\left[P_{\!\odot}^2Q_i^2-3(\vec{P}_{\!\odot}\cdot\vec{Q}_i)^2\right]
	&=&\phm\sum_{i\neq0}\frac{Gm_0m_i}{r_{i0}^5}
	\left[\tilde{v}_0^2r_{i0}^2-3(\tilde{\vec{v}}_0\cdot\vec{x}_{i0})^2\right],
	\\
	\{\{V_\odot,T_0\},T_1\}
	&= \sum_{i\neq0}\frac{G}{Q_i^5}
	\left[(\vec{P}_{\!\odot}\cdot\!\vec{P}_i)Q_i^2
		-3(\vec{P}_{\!\odot}\cdot\vec{Q}_i)(\vec{P}_i\cdot\vec{Q}_i)\right]
	&=&-\sum_{i\neq0}\frac{Gm_0m_i}{r_{i0}^5}
	\left[(\tilde{\vec{v}}_0\cdot\tilde{\vec{v}}_i)r_{i0}^2
		-3(\tilde{\vec{v}}_0\cdot\vec{x}_{i0})(\tilde{\vec{v}}_i\cdot\vec{x}_{i0})
		\right],
	&\qquad &\qquad&\qquad
	\\
	\{\{V_\odot,T_1\},T_1\}
	&= \sum_{i\neq0}\frac{Gm_0}{m_iQ_i^5}
	\left[P_{i}^2Q_i^2-3(\vec{P}_{i}\cdot\vec{Q}_i)^2\right]
	&=&\phm\sum_{i\neq0}\frac{Gm_0m_i}{r_{i0}^5}
	\left[\tilde{v}_{i}^2r_{i0}^2-3(\tilde{\vec{v}}_{i}\cdot\vec{x}_{i0})^2\right],
	\\
	\{\{\Vp,T_1\},T_1\}
	&= \sum_{0<i<j}\frac{Gm_im_j}{Q_{i\!j}^5}
	\left[V_{i\!j}^2Q_{i\!j}^2-3(\vec{V}_{i\!j}\cdot\vec{Q}_{i\!j})^2\right]
	&=&\phm \sum_{0<i<j}\frac{Gm_im_j}{r_{i\!j}^5}
	\left[v_{i\!j}^2r_{i\!j}^2-3(\vec{v}_{i\!j}\cdot\vec{x}_{i\!j})^2\right],
	\\
	\{\{T_0,V_\odot\},V_\odot\}
	&= \frac{1}{m_0}\left(
		\sum_{i\neq0}\frac{Gm_0m_i}{Q_i^3}\vec{Q}_i\right)^2
	&=&\phm\frac{1}{m_0}\left(
		\sum_{i\neq0}\frac{Gm_0m_i}{r_{i0}^3}\vec{x}_{i0}\right)^2,
	\\
	\{\{T_1,V_\odot\},V_\odot\}
	&= \sum_{i\neq0}\frac{G^2m_0^2m_i}{Q_i^4}
	&=&\phm \sum_{i\neq0}\frac{G^2m_0^2m_i}{r_{i0}^4},
	\\
	\{\{T_1,V_\odot\},\Vp\}
	&= \sum_{\substack{i,j>0\\i\neq j}}
		\frac{G^2m_0m_im_{\!j}}{Q_{i\!j}^3Q_i^3}\vec{Q}_{i\!j}\cdot\vec{Q}_i
	&=& \phm \sum_{\substack{i,j>0\\i\neq j}}
		\frac{G^2m_0m_im_{\!j}}{r_{i\!j}^3r_{i0}^3}\vec{x}_{i\!j}\cdot\vec{x}_{i0},
	\\
	\{\{T_1,\Vp\},\Vp\}
	&= \sum_{i\neq0}\frac{1}{m_i}\left(
		\sum_{j\neq0,i}\frac{Gm_im_j}{Q_{i\!j}^3}\vec{Q}_{i\!j}\right)^2
	&=&\phm\sum_{i\neq0}\frac{1}{m_i}\left(
		\sum_{j\neq0,i}\frac{Gm_im_j}{r_{i\!j}^3}\vec{x}_{i\!j}\right)^2
\end{align}
\end{subequations}
with $\vec{V}_i\equiv\vec{P}_i/m_i$ and the negative barycentric Solar momentum
\begin{equation}
\label{eq:barmom}
	\vec{P}_\odot\equiv \sum_{j\neq0}\vec{P}_{\!j} = - m_0\tilde{\vec{v}}_0.
\end{equation}
All other independent nested Poisson brackets of these basic building blocks vanish. For the traditional leapfrog integrators when $H$ is split into kinetic and potential energies ($H = T + V$), 
\begin{equation} 
	\Exp{h \hat{\tilde{H}}} = \Exp{\frac{h}{2} \left(\sub{\hat{V}}{pl} + \sub{\hat{V}}{\odot}\right)} \Exp{h \left(\sub{\hat{T}}{0} + \sub{\hat{T}}{1} \right) }\Exp{\frac{h}{2} \left(\sub{\hat{V}}{pl} + \sub{\hat{V}}{\odot}\right)}
	\qquad\text{or}\qquad
	\Exp{h \hat{\tilde{H}}} = \Exp{\frac{h}{2} \left(\sub{\hat{T}}{0} + \sub{\hat{T}}{1} \right)} \Exp{h \left(\sub{\hat{V}}{pl} + \sub{\hat{V}}{\odot}\right)}\Exp{\frac{h}{2} \left(\sub{\hat{T}}{0} + \sub{\hat{T}}{1} \right)},
\end{equation}
all the Poisson bracket terms above appear in $\tilde{H}$.  

There are several noteworthy properties of these error terms. First, the (only non-vanishing) nested Poisson bracket with two different potential terms $\{\{T_1,V_\odot\},\Vp\}$ is a sum over terms involving only three particles (the Sun and two planets $i,j$), which becomes large only in a close encounter between these three particles \citep{DH16}.  All nested Poisson brackets with only one potential contribution involves two particles (Sun-planet or planet-planet) and becomes large when they are close.  Nested Poisson brackets with two identical potential contributions can be decomposed into sums over terms involving only two particles or only three particles (where `particle' refers to planet or Sun) \citep{DH16}: $\{\{T_0,V_\odot\},V_\odot\}=\{\{T_0,V_\odot\},V_\odot\}_2+\{\{T_0,V_\odot\},V_\odot\}_3$ and analogously for $\{\{T_1,\Vp\},\Vp\}$:
\begin{subequations}
\begin{align}
	\label{eq:Herr:T0V0V0:2}
	\{\{T_0,V_\odot\},V_\odot\}_2
	&= \frac{1}{m_0}
		\sum_{i\neq0} \left(\frac{Gm_0m_i}{Q_i^3}\vec{Q}_i\right)^2
	&=& \phm\sum_{i\neq0}\frac{G^2m_0m_i^2}{r_{i0}^4},
	\\
	\label{eq:Herr:T0V0V0:3}
	\{\{T_0,V_\odot\},V_\odot\}_3
	&= \frac{2}{m_0}\sum_{0<i<j}
		\left(\frac{Gm_0m_i}{Q_{i}^3}\vec{Q}_{i}\right)\cdot
		\left(\frac{Gm_0m_j}{Q_{\!j}^3}\vec{Q}_{\!j}\right)
	&=& 2 \sum_{0<i<j}\frac{G^2m_0m_im_j}{r_{i0}^3r_{\!j0}^3}
		\vec{x}_{i0}\cdot\vec{x}_{\!j0},
	\\
	\label{eq:Herr:T1V1V1:2}
	\{\{T_1,\Vp\},\Vp\}_2
	&= \sum_{i\neq0}\frac{1}{m_i} \sum_{j\neq i}
		\left(\frac{Gm_im_j}{Q_{i\!j}^3}\vec{Q}_{i\!j} \right)^2
	&=&\phm \sum_{0<i<j} \frac{G^2m_im_{\!j}(m_i+m_{\!j})}{r_{i\!j}^4}
	\\
	\label{eq:Herr:T1V1V1:3}
	\{\{T_1,\Vp\},\Vp\}_3
	&= \sum_{i\neq0} \frac{1}{m_i}
		\sum_{j\neq i}\frac{Gm_im_{\!j}}{Q_{i\!j}^3}\vec{Q}_{i\!j}\cdot
		\sum_{k\neq i,j}\frac{Gm_im_{k}}{Q_{ik}^3}\vec{Q}_{ik}
	&=&2 \sum_{i\neq0} \sum_{\substack{0<j<k\\j,k\neq i}}
		\frac{G^2m_im_jm_k}{r_{i\!j}^3r_{ik}^3}
		\vec{x}_{i\!j}\cdot\vec{x}_{ik}
	&=& 2 \sum_{0<i<j<k} G^2m_im_jm_k \sum_{\mathrm{cyclic}(ijk)}
	\frac{\vec{x}_{i\!j}\cdot\vec{x}_{ik}}{r_{i\!j}^3r_{ik}^3}	
\end{align}
\end{subequations}

Second, owing to the vast mass difference between planets and Sun, these terms have different amplitudes. A planet with semi-major axis $a_i$ has typical velocity $\langle \tilde{{v}}_i^2\rangle \sim Gm_0/a_i$ largely independent of the planetary masses. For the Sun, we find from the balance of total momentum, $\tilde{v}_0\lesssim\epsilon\tilde{v}_{i\neq0}$ where $\epsilon=\langle m\rangle/m_0\ll1$ is the typical/dominant planet-to-Sun mass ratio. We have
\begin{subequations} \label{eq:Herr:ampl}
\begin{eqnarray}
	\label{eq:Herr:ampl:1}
	\{\{V_\odot,T_1\},T_1\} \sim
	\{\{T_1,V_\odot\},V_\odot\}
	&\propto& \epsilon^1,
	\\[-0.5ex]
	\{\{T_0,V_\odot\},V_\odot\} \sim
	\{\{T_1,\Vp\},V_\odot\} \sim
	\{\{V_\odot,T_0\},T_1\} \sim
	\{\{\Vp,T_1\},T_1\}
	&\propto& \epsilon^2,
	\\
	\{\{V_\odot,T_0\},T_0\} \sim
	\{\{T_1,\Vp\},\Vp\}
	&\propto& \epsilon^3,
\end{eqnarray}
\end{subequations}
Since all of these terms appear in the second order error Hamiltonian of the leapfrog, its error is dominated by the terms~(\ref{eq:Herr:ampl:1}), making it uninteresting for solving solar system problems. For the WHD integrator
\begin{subequations} \label{eq:err:WHI}
\begin{align}
	\{\{\sub{A}{d},\sub{B}{d}\},\sub{B}{d}\}
		&= \{\{V_\odot,T_0\},T_0\} + \{\{T_1,\Vp\},\Vp\} &\propto& \epsilon^3
	\qquad\text{and}\\
	\{\{\sub{B}{d},\sub{A}{d}\},\sub{A}{d}\}
		&= \{\{T_0,V_\odot\},V_\odot\} + \{\{\Vp,T_1\},T_1\}
		-	\{\{T_1,\Vp\},V_\odot\} - \{\{V_\odot,T_0\},T_1\}  &\propto& \epsilon^2.
\end{align}
\end{subequations}
These scalings can also be derived from\footnote{Ignoring the dynamically irrelevant bulk kinetic energy $\vec{P}_0^2/2M$.} $\sub{A}{d}\propto\epsilon$, $\sub{B}{d}\propto\epsilon^2$, and $\vec{P}_{i\neq0}\propto\epsilon$ such that each Poisson bracket corresponds to a division by $\epsilon$, i.e.\ $\{\{\sub{A}{d},\sub{B}{d}\},\sub{B}{d}\}\propto\epsilon \epsilon^2 \epsilon^2 / \epsilon^2 = \epsilon^3$.  {While our scalings with $\epsilon$ are formally correct, it is conventional (e.g. \cite{Decketal2014}) instead to consider scalings with respect to $\sub{A}{d}$.  In this case $\sub{A}{d}\propto\epsilon^0$, $\sub{B}{d}\propto\epsilon^1$, and $\vec{P}_{i\neq0}\propto\epsilon^0$.  To find conventional scalings, every scaling in this paper should be reduced by a power of $\epsilon$.}

Of the $\mathcal{O}(h^2)$ error terms~(\ref{eqs:Herr:WHD:split}) for the standard leapfrog WHD eliminates those $\propto\epsilon^1$. The presence of the error term $\{\{T_0,V_\odot\},V_\odot\}$, including its $\{\{T_0,V_\odot\},V_\odot\}_2$ part, signifies that the WHD method still incurs errors owing to close Sun-planet encounters, even though it employs a Kepler solver for all of them. This is related to the fact that in the single-planet limit the WHD map does not reduce to a Kepler solver.

The two possible forms of the WHD map,
\begin{equation}\label{eq:WH:maps}
	\phi_h^{\mathrm{BAB}} =
	\Exp{\frac{h}{2}\sub{\op{B}}{d}}
	\Exp{h\sub{\op{A}}{d}}
	\Exp{\frac{h}{2}\sub{\op{B}}{d}}
	\qquad\text{and}\qquad
	\phi_h^{\mathrm{ABA}} =
	\Exp{\frac{h}{2}\sub{\op{A}}{d}}
	\Exp{h\sub{\op{B}}{d}}
	\Exp{\frac{h}{2}\sub{\op{A}}{d}}
\end{equation}
differ in their error Hamiltonian:
\begin{eqnarray}
	\label{eq:WH:Herr:BAB}
	\sub{H}{err}^{\mathrm{BAB}} &=&
	+\frac{h^2}{12}\{\{\sub{B}{d},\sub{A}{d}\},\sub{A}{d}\}
	-\frac{h^2}{24}\{\{\sub{A}{d},\sub{B}{d}\},\sub{B}{d}\}
	+\mathcal{O}(h^4),
	\\
	\label{eq:WH:Herr:ABA}
	\sub{H}{err}^{\mathrm{ABA}} &=&
	-\frac{h^2}{24}\{\{\sub{B}{d},\sub{A}{d}\},\sub{A}{d}\}
	+\frac{h^2}{12}\{\{\sub{A}{d},\sub{B}{d}\},\sub{B}{d}\}
	+\mathcal{O}(h^4).
\end{eqnarray}
The BAB and ABA maps were referred to as $[\mathrm{BA}]^2$ and $[ \mathrm{AB}]^2 $ respectively in \cite{DH16}.  The scalings~(\ref{eq:err:WHI}) suggest to prefer $\phi_h^{\mathrm{ABA}}$ as far as energy or phase space coordinate error, and we can verify numerically that indeed ABA is more accurate than BAB, yet the standard form is $\phi_h^{\mathrm{BAB}}$. This is presumably because it requires only one application of $\Exp{h\sub{\op{A}}{d}}$ per step compared to two for $\phi_h^{\mathrm{ABA}}$. However, when concatenating subsequent steps, the effective number of applications of $\Exp{h\sub{\op{A}}{d}}$ per step can be reduced to one also for $\phi_h^{\mathrm{ABA}}$.    For leapfrog we let $B = V \propto \epsilon$ and $A = T \propto \epsilon$ {(in a barycentric frame)}.  When we use it to solve the Outer Solar System, ABA (also known as drift-kick-drift) has smaller energy error than BAB (also known as kick-drift-kick) again by about a factor 2.  

\begin{figure}
	\includegraphics[width=80mm]{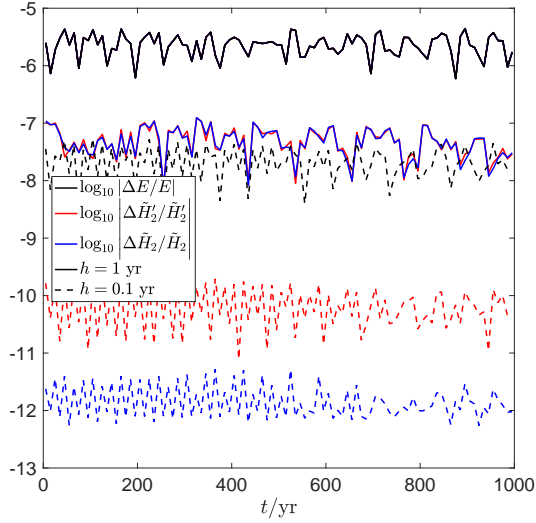}
	\caption{We show that the part of the second order error Hamiltonian which is $\propto \epsilon^3$ only matters if the time step $h$ is small enough.  At larger $h$, fourth order terms in the error Hamiltonian dominate over it.  $\tilde{H}_2$ is the surrogate Hamiltonian truncated beyond the second order in {$h$}. This quantity is conserved up to terms $\mathcal{O}(h^4)$.   $\tilde{H}_2^\prime$ simply subtracts from $\tilde{H}_2$ terms $\propto \epsilon^3$.  We run the outer giants solar system problem with WHD with small step $h = 0.1$ yrs and larger step $h = 1$ yr and the median error of 10 year intervals is plotted.  The error in $\tilde{H}_2$ only differs from $\tilde{H}_2^\prime$ when $h = 0.1$ yr.
	\label{fig:epsilonterms}
  	}
\end{figure}

The fourth order error Hamiltonians for WHD have the terms  
\begin{equation}
\begin{aligned}
&\{\{\{\{\sub{B}{d},\sub{A}{d}\},\sub{A}{d}\},\sub{A}{d}\},\sub{A}{d}\},     
& 
&\{\{\{\{\sub{A}{d},\sub{B}{d}\},\sub{B}{d}\},\sub{B}{d}\},\sub{B}{d}\},  
&
&\{\{\{\{\sub{A}{d},\sub{B}{d}\},\sub{B}{d}\},\sub{B}{d}\},\sub{A}{d}\},
&
&\{\{\{\{\sub{B}{d},\sub{A}{d}\},\sub{A}{d}\},\sub{A}{d}\},\sub{B}{d}\},
&
&\{\{\{\{\sub{A}{d},\sub{B}{d}\},\sub{A}{d}\},\sub{B}{d}\},\sub{A}{d}\}, \\
&\{\{\{\{\sub{B}{d},\sub{A}{d}\},\sub{B}{d}\},\sub{A}{d}\},\sub{B}{d}\}.
\end{aligned}
\end{equation}
$\{\{\{\{\sub{B}{d},\sub{A}{d}\},\sub{A}{d}\},\sub{A}{d}\},\sub{A}{d}\} \propto \epsilon^2$ and will dominate over the second-order term $\{\{\sub{A}{d},\sub{B}{d}\},\sub{B}{d}\}$ if $h$ is large enough.  We show this in Fig. \ref{fig:epsilonterms}.  {If $H_{\mathrm{err},2}$ is the contribution of the error Hamiltonian proportional to $h^2$,} $\tilde{H}_2 = H + H_{\mathrm{err},2}$, while $\tilde{H}_2^\prime$ has the same definition but leaves out terms proportional to $\epsilon^3$.  From \eqref{eq:cons}, if $\Delta \tilde{H}_2$ quantifies the change in $\tilde{H}_2$ from its initial value, $\Delta \tilde{H}_2 \propto h^4$.  We run the outer giant planets problem with a larger $h = 1$ yr and smaller $h = 0.1$ yr using the BAB form of WHD.  {We calculate $E$, $\tilde{H}_2$, and $\tilde{H}_2^\prime$ as a function of discrete time}.  For this problem, $\epsilon \approx 10^{-3}$.  We see that the second order terms $\propto \epsilon^3$ only matter with the smaller $h$.  We have plotted the median error during 10 year intervals to reduce periodic error fluctuations.  

\subsection{An alternative split: WHDS}
\label{sec:WHD:alt}
One issue with the WHD map is that it fails to obtain the exact solution even in the single-planet limit: it keeps the error term $\{\{T_0,V_\odot\},V_\odot\}_2$. This issue can be rectified in democratic heliocentric coordinates by splitting the kinetic energy slightly differently.  Here we discuss this map and derive an error analysis for this map.  The Solar contribution can be split as \citep{LaskarRobutel1995,W06}
\begin{equation}
	\frac{\vec{P}^2_{\!\odot}}{2m_0} \equiv 
	\frac{1}{2m_0} \left(\sum_{i\neq0}\vec{P}_i\right)^2
	=
	\frac{1}{2m_0} \sum_{i\neq0}\vec{P}_i^2 +
	\frac{1}{m_0}\sum_{0<i<j} \vec{P}_i\cdot\vec{P}_j.
\end{equation}
The kinetic energy can be split as $T=\tilde{T}_0+\tilde{T}_1$ with
\begin{equation}
	\tilde{T}_0 = T_0 - \sub{\tilde{T}}{d} =
		\frac{\v{P}_0^2}{2 M} + 
		\frac{1}{2m_0}\left[\left(\sum_{i\neq0}\vec{P}_i\right)^2 -
		\sum_{i\neq0}\vec{P}_i^2\right] =
		\frac{\v{P}_0^2}{2 M} + 
		\frac{1}{m_0}\sum_{0<i<j} \vec{P}_i\cdot\vec{P}_j
	\qquad\text{and}\qquad
	\tilde{T}_1 = T_1 + \sub{\tilde{T}}{d} =
		\sum_{i\neq0}\frac{\vec{P}_i^2}{2\mu_i},
\end{equation}
where the reduced masses are $\mu_i\equiv m_im_0/(m_0+m_i)$, and
\begin{equation}
	\sub{\tilde{T}}{d} \equiv \frac{1}{2m_0} \sum_{i\neq0}\vec{P}_i^2.
\end{equation}
We then have instead of~(\ref{eqs:WHD:A+B})
\begin{subequations} \label{eqs:WH:A+B}
\begin{eqnarray}
	\label{eq:WH:A}
	\sub{\tilde{A}}{d} &=& \sub{A}{d} + \sub{\tilde{T}}{d} = \tilde{T}_1 + V_\odot = 
		\sum_{i\neq0}\frac{\vec{P}_i^2}{2\mu_i} - \frac{G(m_0+m_{i})\mu_i}{Q_i},
	\\ \label{eq:WHD:B}
	\sub{\tilde{B}}{d} &=& \sub{B}{d} - \sub{\tilde{T}}{d} = \tilde{T}_0 + \Vp =
		\frac{\v{P}_0^2}{2 M} + \sum_{0<i<j}  \frac{\vec{P}_i\cdot\vec{P}_j}{m_0}
					- \frac{Gm_im_{\!j}}{Q_{i\!j}}.
\end{eqnarray}
\end{subequations}
Because the Poisson bracket of $\sub{\tilde{T}}{d}$ with the angular or linear momentum is zero, WHDS still conserves these quantities exactly.  The Kepler problems in $\sub{\tilde{A}}{d}$ use the correct gravitating masses $m_0+m_i$ for the unperturbed orbits so this map solves $n=1$ exactly.  There is a small price to pay for these benefits, namely the components of $B$ are no longer in involution such that their order of application matters%
\footnote{
	$\sub{\tilde{B}}{d}$ can be considered the sum over the individual interaction Hamiltonians	\begin{equation}
		\tilde{B}_{i\!j} \equiv \frac{\vec{P}_i\cdot\vec{P}_j}{m_0}
			- \frac{Gm_im_{\!j}}{Q_{i\!j}}.
	\end{equation}
	Curiously, each of these obtains the \emph{repulsive} motion
	\begin{equation}
		\ddot{\vec{Q}}_{i\!j} = \frac{2Gm_im_{\!j}}{m_0} \frac{\vec{Q}_{i\!j}}{Q^3_{i\!j}}
	\end{equation}
	for the relative planetary distance vector and a free motion not of the centre of mass, but of the sum $\vec{Q}_i+\vec{Q}_{\!j}$.
}.
This means applying symplectic correctors from \cite{W06} will no longer be possible, but one may be able to construct other correctors.  The initial value Kepler problems resulting from application of $\Exp{h\op{A}}$ are solved using, for example, universal variables and the Gauss $f$ and $g$ functions as described in \cite{WisdomHernandez2015}.  The program requires using trigonometric functions, square roots, and iterations, resulting in an expensive operation; in fact, the Kepler maps are the {most} significant computational expense of the calculations in this paper. The kick map $\Exp{h\sub{\op{V}}{\!pl}}$ is more expensive than the drift map $\Exp{h \sub{\op{\tilde{T}}}{0}}$.  There are therefore four possible maps for WHDS, but only the following two allow concatenation of either $\Exp{h\op{A}}$ or $\Exp{h\sub{\op{V}}{\!pl}}$ such that on average each of these more expensive maps is applied only once per step:  
\begin{equation}
\label{eq:altsplit}
	\phi_h^{\mathrm{[VTA]^2}} =
	\Exp{\frac{h}{2}\sub{\op{V}}{\!pl}}
	\Exp{\frac{h}{2}\sub{\op{\tilde{T}}}{0}}
	\Exp{h\sub{\op{\tilde{A}}}{d}}
	\Exp{\frac{h}{2}\sub{\op{\tilde{T}}}{0}}
	\Exp{\frac{h}{2}\sub{\op{V}}{\!pl}}
	\qquad\text{and}\qquad
	\phi_h^{\mathrm{[ATV]^2}} =
	\Exp{\frac{h}{2}\sub{\op{\tilde{A}}}{d}}
	\Exp{\frac{h}{2}\sub{\op{\tilde{T}}}{0}}
	\Exp{h\sub{\op{V}}{\!pl}}
	\Exp{\frac{h}{2}\sub{\op{\tilde{T}}}{0}}
	\Exp{\frac{h}{2}\sub{\op{\tilde{A}}}{d}}.
\end{equation}
{Note they are time reversible.}  They have error Hamiltonians
\begin{subequations}
\begin{eqnarray}
\label{eq:splita}
	\sub{H}{err}^{\mathrm{[VTA]^2}} &=&
	+\frac{h^2}{12}\left[
		\{\{\tilde{T}_0,V_\odot\},V_\odot\} +
		\{\{\Vp,T_1\},T_1\} -
		\{V_\odot,\tilde{T}_0\},\tilde{T}_1\} -
		\{\{T_1,V_\odot\},\Vp\}
		\right]
	-\frac{h^2}{24}\left[
		\{\{V_\odot,\tilde{T}_0\},\tilde{T}_0\} +
		\{\{T_1,\Vp\},\Vp\}
		\right]
	+\mathcal{O}(h^4),
	\\
	\label{eq:splitb}
	\sub{H}{err}^{\mathrm{[ATV]^2}} &=&
	-\frac{h^2}{24}\left[
		\{\{\tilde{T}_0,V_\odot\},V_\odot\} +
		\{\{\Vp,\tilde{T}_1\},\tilde{T}_1\} -
		\{V_\odot,\tilde{T}_0\},\tilde{T}_1\} -
		\{\{\tilde{T}_1,V_\odot\},\Vp\}
		\right]
	+\frac{h^2}{12}\left[
		\{\{V_\odot,\tilde{T}_0\},\tilde{T}_0\} +
		\{\{\tilde{T}_1,\Vp\},\Vp\}
		\right]
	\nonumber \\ &&
	-\frac{h^2}{24} \{\{\Vp,\sub{\tilde{T}}{d}\},\sub{\tilde{T}}{d}\}
	+\frac{h^2}{12}\left[
		\{\{\Vp,\sub{\tilde{T}}{d}\},\tilde{T}_1\} \,{-}\,
		\{\{\sub{\tilde{T}}{{0}},V_\odot\},\Vp\} -
		\{\{\sub{\tilde{T}}{d},\Vp\},\Vp\}
		\right]	+\mathcal{O}(h^4).
\end{eqnarray}
\end{subequations}
At second order $\sub{H}{err}^{\mathrm{[VTA]^2}}$ is identical to $\sub{H}{err}^{\mathrm{BAB}}$ from WHD except for the replacements
\begin{subequations}
	\label{eqs:VTATV:BAB}
\begin{eqnarray}
	\{\{T_0,V_\odot\},V_\odot\} &\to \{\{\tilde{T}_0,V_\odot\},V_\odot\} = &
		\{\{T_0,V_\odot\},V_\odot\}_3,
	\\
	\{\{V_\odot,T_0\},T_0\} &\to \{\{V_\odot,\tilde{T}_0\},\tilde{T}_0\} \;=&
	\sum_{i\neq0}\frac{Gm_i}{m_0Q_i^5}
	\left[(\vec{P}_{\!\odot}-\vec{P}_i)^2Q_i^2-3\left((\vec{P}_{\!\odot}-\vec{P}_i)
		\cdot\vec{Q}_i\right)^2\right],
	\\
	\{\{V_\odot,T_0\},T_1\} &\to \{\{V_\odot,\tilde{T}_0\},\tilde{T}_1\} \;=&
	\sum_{i\neq0}\frac{Gm_i}{\mu_iQ_i^5}
	\left[\left([\vec{P}_{\!\odot}-\vec{P}_i]\cdot\!\vec{P}_i\right)Q_i^2
		-3\left([\vec{P}_{\!\odot}-\vec{P}_i]\cdot\vec{Q}_i\right)
			(\vec{P}_i\cdot\vec{Q}_i)\right].
\end{eqnarray}
\end{subequations}
Most significantly, the error term $\{\{T_0,V_\odot\},V_\odot\}_2$ has been eliminated, which is a direct consequence of the correctness of the map in the single-planet limit.
The other differences appear to be relatively minor.  The difference between $\sub{H}{err}^{\mathrm{[ATV]^2}}$ and $\sub{H}{err}^{\mathrm{ABA}}$ is larger: in addition to the replacements~(\ref{eqs:VTATV:BAB}) also the remaining three terms are replaced and additional terms appear.

Instead of equations~(\ref{eqs:Herr:WHD:split}), we must consider the terms (note that sums never include the index 0, even if not explicitly stated)
\begin{subequations}
	\label{eqs:Herr:WH:split}
\begin{align}
%
%
%
%
	\{\{V_\odot,\tilde{T}_0\},\tilde{T}_0\}
	&= \sum_{i}\frac{Gm_i}{m_0Q_i^5}
	\left[(\vec{P}_{\!\odot}-\vec{P}_i)^2Q_i^2-3\left((\vec{P}_{\!\odot}-\vec{P}_i)
		\cdot\vec{Q}_i\right)^2\right]
	&\propto&\epsilon^3,
	\\
	\{\{V_\odot,\tilde{T}_0\},\tilde{T}_1\}
	&= \sum_{i}\frac{Gm_i}{\mu_iQ_i^5}
	\left[\left([\vec{P}_{\!\odot}-\vec{P}_i]\cdot\!\vec{P}_i\right)Q_i^2
		-3\left([\vec{P}_{\!\odot}-\vec{P}_i]\cdot\vec{Q}_i\right)
			(\vec{P}_i\cdot\vec{Q}_i)\right]
	&\propto&\epsilon^2,
	\\
	\{\{V_\odot,\tilde{T}_1\},\tilde{T}_1\}
	&= \sum_{i}\frac{G(m_0+m_i)}{\mu_iQ_i^5}
	\left[P_{i}^2Q_i^2-3(\vec{P}_{i}\cdot\vec{Q}_i)^2\right]
	&\propto&\epsilon^1,
	\\
	\{\{\Vp,\tilde{T}_0\},\tilde{T}_0\}
	&= \sum_{i<j}\frac{Gm_im_{\!j}}{m_0^2Q_{i\!{j}}^5}
		\left[P_{i\!j}^2Q_{i\!j}^2-3(\vec{P}_{i\!j}\cdot\vec{Q}_{i\!j})^2\right]
	&\propto&\epsilon^4,
	\\
	\{\{\Vp,\tilde{T}_0\},\tilde{T}_1\}
	&= -\sum_{i<j}\frac{Gm_im_{\!j}}{m_0Q_{i\!{j}}^5}
		\left[\vec{P}_{i\!j}\cdot\tilde{\vec{w}}_{i\!j}Q_{i\!j}^2
			-3(\vec{P}_{i\!j}\cdot\vec{Q}_{i\!j})
			  (\tilde{\vec{w}}_{i\!j}\cdot\vec{Q}_{i\!j})\right]
	&\propto&\epsilon^3,
	\\
	\{\{\Vp,\tilde{T}_1\},\tilde{T}_1\}
	&= \sum_{i<j}\frac{Gm_im_j}{Q_{i\!j}^5}
	\left[\tilde{w}_{i\!j}^2Q_{i\!j}^2-
		3(\tilde{\vec{w}}_{i\!j}\cdot\vec{Q}_{i\!j})^2\right]
	&\propto&\epsilon^2,
	\\
%
%
	\{\{\tilde{T}_0,V_\odot\},V_\odot\}
	&= \{\{T_0,V_\odot\},V_\odot\}_3
	 = 2 \sum_{i<j}\frac{G^2m_0m_im_{\!j}}{Q_i^3Q_{\!j}^3}
		\vec{Q}_i\cdot\vec{Q}_{\!j}
	&\propto&\epsilon^2,
	\\
	\{\{\tilde{T}_0,V_\odot\},\Vp\}
	&= \sum_{i} \sum_{j,k\neq i}\frac{G^{{2}}m_im_{\!j}m_k}{Q_{i\!j}^3Q_{k}^3}
		\vec{Q}_{i\!j}\cdot\vec{Q}_{k}
	&\propto&\epsilon^3,
	\\
	\{\{\tilde{T}_0,\Vp\},\Vp\}
	&=-\frac{1}{m_0}\sum_i\left(\sum_{j\neq i}\frac{Gm_im_{\!j}}{Q_{i\!j}^3}
		\vec{Q}_{i\!j}\right)^2
	&\propto&\epsilon^4,
	\\
	\{\{\tilde{T}_1,V_\odot\},V_\odot\}
	&= \sum_{i}\frac{G^2m_0(m_0+m_i)m_i}{Q_i^4}
	&\propto&\epsilon^1,
	\\
	\{\{\tilde{T}_1,V_\odot\},\Vp\}
	&= \sum_{i\neq j}\frac{G^2(m_0+m_i)m_im_{\!j}}{Q_i^3Q_{i\!j}^3}
		\vec{Q}_i\cdot\vec{Q}_{i\!j}
	&\propto&\epsilon^2,
	\\
	\{\{\tilde{T}_1,\Vp\},\Vp\}
	&= \sum_i\frac{1}{\mu_i}\left(
		\sum_{j\neq i}\frac{Gm_im_j}{Q_{i\!j}^{{3}}}\vec{Q}_{i\!j}\right)^2
	&\propto&\epsilon^3
\end{align}
\end{subequations}
with the enhanced barycentric velocities $\tilde{\vec{w}}_i\equiv\vec{P}_i/\mu_i$.
\subsection{A map in Canonical Heliocentric coordinates: CH}
\label{sec:chc}
We now discuss a symplectic map that uses the Canonical Heliocentric coordinates from \cite{LaskarRobutel1995} but{, compared to them, we restore the full Hamiltonian, as described in Sec. \ref{sec:chcoord}, before splitting into $A$ and $B$}.  The difference between these coordinates with the Democratic Heliocentric coordinates is that the $\v{P}_i$ are now inertial, not necessarily barycentric, momenta and $\v{Q}_0 = \v{x}_0$.  We carry out error analysis for general inertial frames.  We need functions  
\begin{subequations}
\begin{eqnarray}
\sub{A}{ch} &=& \sub{T}{pl} + V_\odot = 
		\sum_{i\neq0}\frac{\vec{P}_i^2}{2m_i} - \frac{Gm_0m_i}{Q_i}, \qquad \text{ and}
		\\
\sub{B}{ch} &=& T_{\mathrm{cm}} + \sub{{T}}{ch,1} + \Tc + \Vp =
		\frac{\v{P}_0^2}{2 m_0} +
		\frac{1}{2m_0}\left(\sum_{i\neq0}\vec{P}_i\right)^2
		- \frac{\v{P}_0}{m_0} \cdot \sum_{i \ne 0} \v{P}_i
		- \sum_{0<i<j}\frac{Gm_im_{\!j}}{Q_{i\!j}}.
\end{eqnarray}
\end{subequations}
The four functions in $\sub{B}{ch}$ are proportional to $\epsilon^2$ and are in involution amongst themselves.  These four functions and $\sub{T}{pl}$ and $\sub{V}{\odot}$ have vanishing Poisson bracket with the linear and angular momenta so these are conserved exactly by CH.  {The purpose of the splitting into $\sub{A}{ch}$ and $\sub{B}{ch}$ is that the two-body problems of $\sub{A}{ch}$ approximate the Sun-planet Kepler problems.  This approximation worsens with increasing ${P}_0$.  In fact, for large enough ${P}_0$, the $\sub{A}{ch}$ orbits are unbound and hyperbolic, which is wrong.}  The conserved quantity $\v{R}$ takes form $\v{R} =  M \v{Q}_0 + \sum_{i \ne 0} m_i \v{Q}_i - t \v{P}_0$.  
\begin{equation}
\{\v{R},A_{\mathrm{ch}}\} = \v{P}_0 - \v{p}_0 \qquad \text{and} \qquad \{\v{R},B_{\mathrm{ch}}\} = \v{p}_0 ,
\end{equation}
implying $\{\v{R},H_{\mathrm{err}}\} \ne 0$, which means the center of mass position is not accurately calculated with this map.  This equation will reappear for a different map in \eqref{eq:WHI:Xcm}.  The center of mass error will not matter for some purposes, but it can be corrected easily simply by correctly updating $\v{Q}_0$.   

The error terms to second order are still listed in eq. \eqref{eqs:Herr:WHD:split}, but we have three additional terms when $\v{P}_0 \ne 0$: 
\begin{subequations}
	\label{eqs:Herr:CH}
\begin{align}
	\{\{V_\odot,{T}_{\mathrm{ch}}\},\sub{{T}}{ch,1}\}
	&= -\sum_{i \ne 0}\frac{Gm_i}{m_0 Q_i^5}
	\left[Q_i^2 (\v{P}_0 \cdot \v{P}_\odot) - 3 (\v{P}_{\odot}  \cdot \v{Q}_i)(\v{P}_0\cdot \v{Q}_i )  \right],
	&
	\\
	\{\{V_\odot,{T}_{\mathrm{ch}}\},\sub{T}{pl}\}
	&= -\sum_{i \ne 0}\frac{G}{Q_i^5}
	\left[Q_i^2 (\v{P}_0 \cdot \v{P}_i) - 3 (\v{P}_i  \cdot \v{Q}_i)(\v{P}_0  \cdot \v{Q}_i)  \right],
	&
	\\
	\{\{V_\odot,{T}_{\mathrm{ch}}\},{T}_{\mathrm{ch}}\}
	&= \sum_{i \ne 0}\frac{Gm_i}{m_0 Q_i^5}
	\left[Q_i^2 P_0^2 - 3 (\v{P}_0 \cdot \v{Q}_i )^2  \right].
	&
	\\
\end{align}
\end{subequations}
{Note the scaling with $\epsilon$ is indeterminate due to $\v{P}_0$.}  The nested Poisson brackets at second order are
\begin{subequations} \label{eq:err:CH}
\begin{align}
	\{\{\sub{A}{ch},\sub{B}{ch}\},\sub{B}{ch}\}
		&= \{\{V_\odot,T_0\},T_0\} + \{\{T_1,\Vp\},\Vp\} +  2 \{\{V_\odot,{T}_{\mathrm{ch}}\},{T}_0\} + \{\{V_\odot,{T}_{\mathrm{ch}}\},{T}_{\mathrm{ch}}\} &\propto& \epsilon^3
	\qquad\text{and}\\
	\{\{\sub{B}{ch},\sub{A}{ch}\},\sub{A}{ch}\}
		&= \{\{T_0,V_\odot\},V_\odot\} + \{\{\Vp,T_1\},T_1\}
		-	\{\{T_1,\Vp\},V_\odot\} - \{\{V_\odot,T_0\},T_1\}  - \{\{V_\odot,{T}_{\mathrm{ch}}\},{T}_1\}&\propto& \epsilon^2,
\end{align}
\end{subequations}
and the form of the error Hamiltonians is still \eqref{eq:WH:Herr:BAB} and \eqref{eq:WH:Herr:ABA}.  {They imply the error in energy $\propto h^2$ while the error in $\tilde{H}_2 \propto h^4$.  The scalings with $\epsilon$ only work in a barycentric frame.}This map again is unable to solve the $n=1$ problem exactly due to $\{\{T_0,V_\odot\},V_\odot\}$.
The energy error is minimized in a barycentric frame.  To study the difference of CH with WHD, write $\sub{A}{ch} = \sub{A}{d} + \Delta$ and $\sub{B}{ch} = \sub{B}{d} - \Delta $.  $\Delta$ is just given by \eqref{eq:Delta}, which will be discussed in Sec. \ref{sec:WHI}.  Although it is written in different coordinates, the CH map is identical, neglecting roundoff error, to a new Wisdom-Holman method, WHI, which we derive in that section.  {In a barycentric frame, $\Delta = 0$ and the CH map is closely related to the WHD map (Section \ref{sec:relation})}.  

\begin{figure}
	\includegraphics[width=80mm]{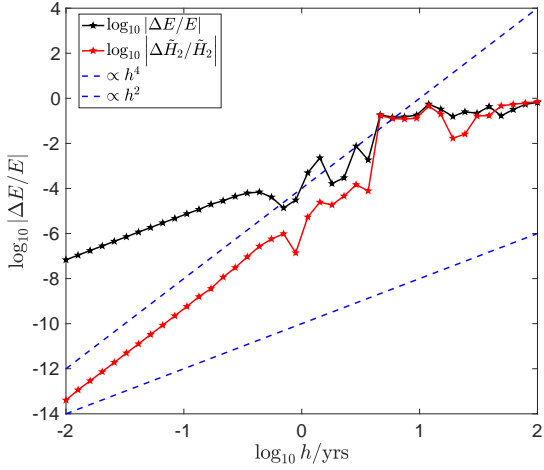}
	\caption{We test the error in energy and in the second order error Hamiltonian of the Canonical Heliocentric map.  As expected, they scale as $\propto h^2$ and $\propto h^4$, respectively if $h$ is not large.  We integrate the outer giant planets in a non-barycentric frame for 1000 yrs. 	
	\label{fig:canhel}
  	}
\end{figure}

Our goal is to test the error terms in \eqref{eq:err:CH}.  We use the CH BAB map and run the outer solar system for $t = 1000$ yrs with various $h$ in a non-barycentric frame.  In the center of mass frame, the outer solar system has an RMS momentum vector of $(1.34,0.789,0.369) \times 10^{-3} M_\odot$ au/yr; we use initial conditions in \cite{HLW06}.  We shift to a frame with total momentum vectors $(2.26,-0.891,0.448) \times 10^{-1} M_\odot$ au/yr: each component is more than 100 times larger in magnitude than its corresponding RMS momentum component.  The error in energy and surrogate Hamiltonian up to second order is shown in Fig. \ref{fig:canhel} and, as expected, scale as $h^2$ and $h^4$ respectively as long as $h$ is not large.

Like in Section \ref{sec:WHD:alt}, if we move the piece 
\begin{equation}
\label{eq:movei}
\tilde{T}_{\mathrm{c}} \equiv \frac{1}{2 m_0} \sum_{i \ne 0} \v{P}_i^2
\end{equation}
from $\sub{B}{ch}$ to $\sub{A}{ch}$, the $n = 1$ {problem} is solved exactly.  In fact, we will obtain the same map as that of Section \ref{sec:WHI:alt}.  

\subsection{The Wisdom-Holman map in Jacobi coordinates: WHJ}
\label{sec:WHJ}
This method was originally used \citep{WH91} to study Pluto's orbit.  Here we provide and test new error analysis for this map for general gravitating masses and analyze the dependences on $\epsilon$ of the method.
Using eq. \eqref{eq:jacT}, 
\begin{subequations}
\begin{eqnarray}
\sub{A}{J} &=& \sub{T}{J} + V_1 = \sum_{i \ne 0} \frac{\v{s}_i^2}{2 m_i^\prime} - \frac{G m_i^\prime \mathcal{M}_i}{u_i},
\\
\sub{B}{J} &=&  \sub{T}{J0} -V_1 + V_2 = \frac{\v{s}_0^2}{2 m_0^\prime}  + \sum_{i \ne 0} \frac{G m_i^\prime \mathcal{M}_i}{u_i} - \sum_{i < j} \frac{G m_i m_j}{r_{i j}},
\end{eqnarray}
\end{subequations}
where $\mathcal{M}_i$ is the gravitating mass of the two-body problem.  Two forms for the gravitating mass are the original form $\mathcal{M}_i = m_0 M_i/M_{i - 1}$, and $\mathcal{M}_i = M_i$ used by \cite{RT15}.  We recover the $n=1$ Hamiltonian \eqref{eq:neq1} as long as $\mathcal{M}_1 = m_0 + m_1$ which is the case for both these choices.  A third choice for the gravitating mass which satisfies the $n = 1$ requirement is $\mathcal{M}_i = m_0 + m_i$ \citep{W17}.  The physical interpretation of the latter two gravitating masses is clearest: the second one gives the correct monopole force; the gravitating masses are $M_i$.  In the final choice, the masses entering the period equation for the unperturbed orbits are $m_i + m_0$.  

Note this map has mixed Jacobi and Cartesian coordinates, so its implementation will not be as straightforward.  One strategy is to carry out calculations in Jacobi coordinates, except those involving $V_2$.  For $V_2$ we calculate accelerations in Cartesian coordinates and convert these to accelerations in Jacobi coordinates using \eqref{eq:tojac}.  {In long term calculations, performing these coordinate transformations with care \citep{RT15} will ensure the energy and angular momentum roundoff error grows at the theoretical minimum rate of time$^{1/2}$. \footnote{\cite{Hernandez2016} has a typo in the introduction and states that the method HB15 follows this theoretical minimum, also known as Brouwer's Law.  The rest of the paper correctly shows the opposite is true.}}  

\begin{figure}
	\includegraphics[width=80mm]{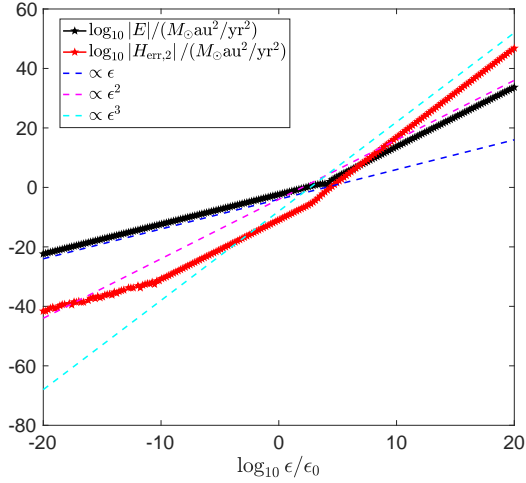}
	\caption{We test the $\epsilon$ dependence of the energy and the second order piece of the error Hamiltonian for the Wisdom-Holman method in Jacobi coordinates.  $H_{\mathrm{err,2}}$ scales initially as $\epsilon$ due to finite precision effects.  Then it scales as $\epsilon^2$ and then as $\epsilon^3$, as expected.  We use the initial conditions of the outer giant planets and modify $\epsilon$ by scaling the mass of the planets: $\epsilon/\epsilon_0 = 1$ corresponds to the usual solar system.  {Scalings in the literature are typically defined with respect to $\sub{A}{J}$ and depend on one fewer power of $\epsilon$.}  
	\label{fig:epsilonwhj}
  	}
\end{figure}

In Jacobi coordinates $\v{R} = \v{u}_0 - t \v{s}_0$ and    
\begin{equation}
	\label{eq:WHJ:Xcm}
	\{\vec{R},\sub{A}{J}\} = 0
	\quad\text{and}\quad
	\{\vec{R},\sub{B}{J}\} = \v{s}_0.
\end{equation}
So $d\v{R}/dt = \{\v{R},\tilde{H}\} + {\partial \v{R}}/{\partial t} = \v{s}_0 - \v{s}_0 =  0$: WHJ integrates correctly the center of mass motion.  Because $\v{u}_0$ is cyclic, we ignore $ \sub{T}{J0}$ in what follows.  We find  
\begin{equation}
\sub{B}{J} = \sum_{i \ne 0}^n {G m_i m_0} \left( \frac{\mathcal{M}_i M_{i - 1}}{m_0 M_i}\frac{1}{u_i} - \frac{1}{r_{i0}} \right) - \sum_{0<i<j} \frac{G m_i m_j}{r_{i j}} = - \sum_{0 < i < j} \left( - \frac{G m_i m_j \v{u}_i \cdot \v{u}_j}{u_j^3} + \frac{G m_i m_j}{u_{i j}} \right) + \mathcal O(\epsilon^3) \propto \epsilon^2:
\end{equation}
$\sub{B}{J} \propto \epsilon^2$ and $\sub{A}{J} \propto \epsilon$, as expected.  The scalars $\sub{T}{J}$, $V_1$, and $\sub{T}{J0}$ have zero Poisson bracket with the angular and linear momentum so WHJ conserves momenta exactly.  To compute $H_{\mathrm{err},2}$, we need
\begin{subequations}
	\label{eqs:Herr:WH:jac}
\begin{align}
\label{eq:tt}
	\{\{V_1,\sub{T}{J}\},\sub{T}{J}\}
	&= \sum_{i \ne 0}\frac{G \mathcal{M}_i}{m_i^\prime u_i^5}
	\left[u_i^2 s_i^2 - 3 (\v{u}_i \cdot \v{s}_i)^2 \right]
	&\propto&\epsilon,
	\\
	\{\{V_2, \sub{T}{J} \}, \sub{T}{J}\}
	&= \sum_{i, j < i} \frac{G m_i m_j}{r_{i j}^5}
	\left[r_{i j}^2 v_{i j}^2 - 3 (\v{x}_{i j} \cdot \v{v}_{i j})^2 \right]
	&\propto&\epsilon,
	\\
	\{\{\sub{T}{J}, V_1\}, V_1\}
	&= \sum_{i \ne 0} \frac{G^2 \mathcal{M}_i^2 m_i^\prime}{u_i^4}
	&\propto&\epsilon,
	\\
	\{\{\sub{T}{J}, V_2 \}, V_2 \}
	&= \sum_{i} \frac{1}{m_i} 
	\left( \sum_{j \ne i} \frac{G m_i m_j}{r_{i j}^3} \v{x}_{i j} \right)^2
	&\propto&\epsilon,
	\\
	\label{eq:mix}
	\{\{\sub{T}{J}, V_2\}, V_1 \}
	&= \sum_{i \ne 0, j \ne i} \frac{G^2 m_i^\prime \mathcal{M}_i m_j}{u_i^3 r_{i j}^3} \v{u}_i \cdot \v{x}_{i j}	 - 
	\sum_{k \ne 0} \sum_{i = 0}^{k-1} \sum_{j \ne i} \frac{G^2 m_k \mathcal{M}_k m_i m_j}{M_k u_k^3 r_{i j}^3} \v{u}_k \cdot \v{x}_{i j}
	&\propto&\epsilon.
	\end{align}
\end{subequations}
We have written the nested Poisson brackets in terms of Jacobi, Cartesian, or both Jacobi and Cartesian coordinates based on convenience.  None of these Poisson brackets or the others that appear at higher orders depend on the velocity frame of reference.  To write the error Hamiltonians, we have
\begin{eqnarray}
\label{eq:sub1}
	\sub{H}{err,2}^{\mathrm{BAB}} &=& \frac{h^2}{24}
	\Bigg[2 \bigg(\{\{V_2,\sub{T}{J}\},\sub{T}{J}\} - \{\{V_1,\sub{T}{J}\},\sub{T}{J}\}   \bigg) +
	\bigg(\{\{ \sub{T}{J}, V_1\},V_1\} -  \{\{ \sub{T}{J}, V_2\},V_2\}  \bigg)\Bigg],
	\\
	\sub{H}{err,2}^{\mathrm{ABA}} &=& \frac{h^2}{24}
	\Bigg[ \bigg(\{\{V_1,\sub{T}{J}\},\sub{T}{J}\} - \{\{V_2,\sub{T}{J}\},\sub{T}{J}\} \bigg) + \bigg(\{\{ \sub{T}{J}, V_1\},V_1\} + 2 \{\{ \sub{T}{J}, V_2\},V_2\} - 3 \{\{ \sub{T}{J}, V_2\},V_1\}  \bigg)  \Bigg].
\end{eqnarray}
The $\mathcal{O}(\epsilon)$ terms in parenthesis cancel, leaving us with $H_{\mathrm{err},2} \propto \epsilon^2$.  To verify this scaling, we calculate the energy and $H_{\mathrm{err,2}}$ for the BAB map for the {initial conditions of the} outer giant planets problem, with planetary masses scaled by $\epsilon/\epsilon_0$ ($\epsilon/\epsilon_0 = 1$ is the usual Solar System).  We use a barycentric frame and $\mathcal{M}_i = M_i$. The result is shown in Fig. \ref{fig:epsilonwhj}.
As expected, the energy scales initially as $\epsilon$ and finally as $\epsilon^2$.  $H_{\mathrm{err,2}}$ has expected $\propto \epsilon^2$ and $\propto \epsilon^3$ behaviors, but there is an unexpected $\propto \epsilon$ behavior.  A closer look reveals this behavior is due to machine precision error.  When $\log_{10} \epsilon/ \epsilon_0 \lesssim -11$, the four Poisson bracket terms of \eqref{eq:sub1} are equal to 15 digits of precision, nearly the machine precision in this case.  Thus, the subtractions in \eqref{eq:sub1} are dominated by finite precision effects.  

\begin{figure}
	\includegraphics[width=80mm]{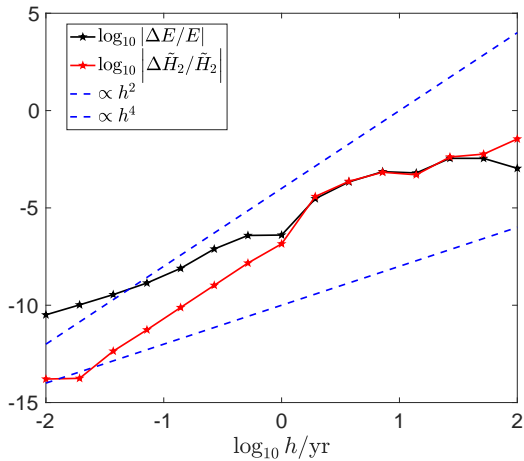}
	\caption{We verify that we have computed correctly the second order error Hamiltonian of the Wisdom-Holman method in Jacobi coordinates and implemented the algorithm correctly.  The energy error and $\tilde{H}_2$ error scale as $h^2$ and $h^4$, respectively, as long as $h$ is not large.  We have run the outer giant planets problem with WHJ for 1000 yrs. 
	\label{fig:htildwhj}
  	}
\end{figure}

Next we verify that the energy error scales as $h^2$ while the error in $\tilde{H}_2 = E + H_{\mathrm{err,2}}$ scales as $h^4$.  We use the same problem with fixed $\epsilon/\epsilon_0 = 1$, run for $t = 1000$ yrs, and use various time steps.  Our verification is shown in Fig. \ref{fig:htildwhj}.
At large $h$, the expected scaling is lost and the convergence of the error Hamiltonian is in danger.  

\begin{figure}
	\includegraphics[width=80mm]{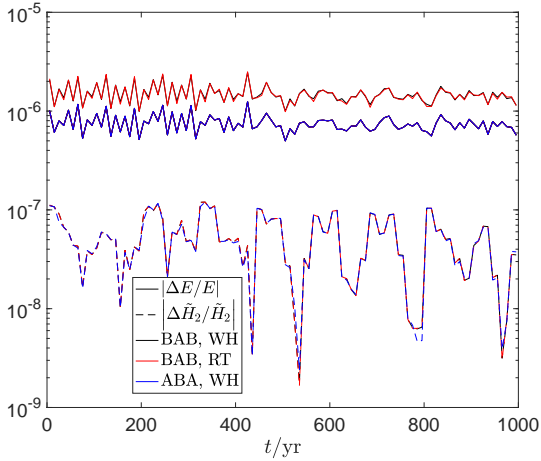}
	\caption{We first test how changing the gravitating mass in the Wisdom-Holman method in Jacobi coordinates, using the \citeauthor{RT15} and \citeauthor{WH91} choices, affects integrator performance.  We find the difference is negligible.  Next we verify the energy error is smaller when using the ABA version of the integrator vs the BAB version.  However, there is little difference in the second order Hamiltonian error.  We integrate the outer giant planets with step $h = 1$ yr and reduce error oscillations by plotting the median errors every 10 years.  
\label{fig:whjcompare}
  	}
\end{figure}

Finally, we test the same problem, with $h = 1$ yr, in Fig. \ref{fig:whjcompare}, to verify that ABA is better than BAB by approximately factor two in energy error (the solid blue and black curves).  The dashed blue and black curves show that the error in $\tilde{H}_2$, which equals a term we have not derived that is proportional to $h^4$, however, does not change.  We compare the gravitating mass choices of \citeauthor{RT15} and \citeauthor{WH91} in the black and red curves: their behaviour is essentially the same, which was also reported by \cite{RT15}.


\subsection{The Wisdom-Holman map in inertial Coordinates: WHI}
\label{sec:WHI}

The simplest possibility for a symplectic map is to use ordinary inertial Cartesian coordinates and split the kinetic energy into the Solar and planetary contribution and combine the latter with the Sun-planet gravitational interaction to obtain a Hamiltonian of $n$ decoupled Kepler orbits. Despite the simplicity of this map, we have not found it discussed in the literature, perhaps because it loses effectiveness in a non-barycentric frame.  We will show it is equivalent to the CH map and it reduces to the WHD map in the barycentric frame.

We split the kinetic energy into
\begin{equation}
	T_{\odot} = \frac{\vec{p}_0^2}{2m_0}
	\qquad\text{and}\qquad
	\Tp  = \sum_{i\neq0}\frac{\vec{p}_i^2}{2m_i}.
\end{equation}
such that instead of equations~(\ref{eqs:WHD:A+B}) we have
\begin{equation} \label{eqs:WHI:A+B}
	\sub{A}{i} = \Tp + V_\odot =
		\sum_{i\neq0}\frac{\vec{p}_i^2}{2m_i} - \frac{Gm_0m_i}{r_{i 0}}
	\qquad\text{and}\qquad
	\sub{B}{i} = T_\odot + \Vp = \frac{\vec{p}_0^2}{2m_0}
		- \sum_{0<i<j}\frac{Gm_im_{\!j}}{r_{i j}}.
\end{equation}
As discussed in Section \ref{sec:chc}, as ${p}_\mathrm{tot}$ increases, the $\sub{A}{i}$ functions no longer resemble {the physical two-body} Kepler problems {we are interested in solving} and the map becomes meaningless.  $\sub{A}{i}$ and $\sub{B}{i}$ are integrable.  The Poisson bracket of $T_\odot$ with the momenta is zero, and all other kinetic and potential term Poisson brackets with momenta are zero as we found in Section \ref{sec:chc}; thus, WHI conserves angular and linear momenta exactly.  $\sub{B}{i}$ generates a kick of the planets due to their mutual gravity and a drift of the Sun (a drift updates the position assuming a constant velocity), while $\sub{A}{i}$ generates Keplerian orbits of the planets around the stationary Sun, but with gravitating mass $m_0$ instead of $m_0+m_i$, plus an update to the Solar momentum such that the total momentum is conserved.  The $\mathcal{O}(h^2)$ error terms for the resulting WHI map are given by  ~(\ref{eqs:Herr:WHD:split}), with substitutions $T_0 \rightarrow T_\odot$, $T_1 \rightarrow \Tp$, and $\tilde{\vec{v}}_i \rightarrow \vec{v}_i$.  So the error Hamiltonian still has two-body error terms and $n=1$ is again not solved exactly.   

\subsubsection{Relation to the Wisdom-Holman map in democratic heliocentric coordinates}
\label{sec:relation}
We explore the relationship of the WHI integrator with WHD from Section \ref{sec:WHD}.  The mapped phase space from WHI, transformed to Canonical Heliocentric coordinates, is equal to the mapped phase space from the Canonical Heliocentric map (if we neglect roundoff error).  

In the barycentric frame, $\sub{A}{i}=\sub{A}{d}$, $\sub{B}{i}=\sub{B}{d}$, $\{\{\sub{A}{i},\sub{B}{i}\},\sub{B}{i}\}=\{\{\sub{A}{d},\sub{B}{d}\},\sub{B}{d}\}$, and $\{\{\sub{B}{i},\sub{A}{i}\},\sub{A}{i}\}=\{\{\sub{B}{d},\sub{A}{d}\},\sub{A}{d}\}$. Yet, WHI and WHD are different, since WHD exactly conserves the constant of motion
\begin{equation}
	\vec{R}\equiv \sum_i m_i\vec{x}_i - t\vec{p}_i
	= M\vec{Q}_0 - t\vec{P}_0
\end{equation}
while WHI does not, i.e. it does not exactly integrate the centre-of-mass position.  We find
\begin{equation}
	\label{eq:WHI:Xcm}
	\{\vec{R},\sub{A}{i}\} = \vec{P}_0 - {\vec{p}_0}, 
	\quad\text{and}\quad
	\{\vec{R},\sub{B}{i}\} = {\vec{p}_0},
\end{equation}
so that ${R}$ is not constant.  The relations \eqref{eq:WHI:Xcm} are the same for the CH map, substituting $\sub{A}{i} \rightarrow \sub{A}{ch}$ and $\sub{B}{i} \rightarrow \sub{B}{ch}$.  In contrast, for the WHD map
\begin{equation}
	\label{eq:WHD:Xcm}
	\{\vec{R},\sub{A}{d}\} = 0
	\quad\text{and}\quad
	\{\vec{R},\sub{B}{d}\} = \v{P}_0,
\end{equation}
which leads to $d\v{R}/dt = \v{P}_0 - \v{P}_0 =  0$, for the continuous problem represented by $\tilde{H} = H + H_{\mathrm{err}}$.  The apparent contradiction occurs because the \emph{functions} $\sub{A}{i}=\sub{A}{d}+\Delta\neq\sub{A}{d}$ and $\sub{B}{i}=\sub{B}{d}-\Delta\neq\sub{B}{d}$ with difference
\begin{equation}
\label{eq:Delta}
	\Delta
	 =  \frac{M-m_0}{2M^2}\vec{P}_0^2 + \frac{1}{M}\vec{P}_0\cdot\sum_{i\neq0}\vec{P}_i
	 = \frac{M+m_0}{2M^2}\sub{\vec{p}}{tot}^2 - \frac{1}{M}\sub{\vec{p}}{tot}\cdot\vec{p}_0,
\end{equation}
which vanishes in the barycentric frame and satisfies $\{T_0,\Delta\}=\{T_1,\Delta\}=\{\Vp,\Delta\}=\{\sub{B}{d},\Delta\}={\{\vec{P}_{i\neq0},\Delta\}}=0$.
\begin{equation}
	\{\sub{A}{d},\Delta\} = \{V_\odot,\Delta\} = \frac{\vec{P}_0}{M}\cdot
		\sum_{i\neq0}\frac{Gm_0m_i}{Q_i^3}\vec{Q}_i =
		\vec{F}_\odot\cdot\sub{\vec{v}}{\!cm}
\end{equation}
with $\vec{F}_\odot$ the total planetary force pulling the Sun, and
\begin{equation}
	\{\vec{R},\Delta\} = \vec{P}_0 - {\vec{p}_0}.
\end{equation}
Thus, WHI differs from WHD in the way the kinetic energy is split such that for $\sub{\vec{p}}{tot}=0$ it agrees in value but not in the derivative. The difference between the second-order error terms can be expressed from
\begin{subequations}
	\label{eqs:h2:diff}
\begin{eqnarray}
	\{\{\sub{A}{i},\sub{B}{i}\},\sub{B}{i}\} -
	\{\{\sub{A}{d},\sub{B}{d}\},\sub{B}{d}\}
	&=& \{\{\sub{A}{d},\Delta\},\Delta-2\sub{B}{d}\}
	\\
	\{\{\sub{B}{i},\sub{A}{i}\},\sub{A}{i}\} -
	\{\{\sub{B}{d},\sub{A}{d}\},\sub{A}{d}\}
	&=& \{\{\sub{A}{d},\Delta\},\sub{A}{d}+\Delta-\sub{B}{d}\},
\end{eqnarray}
\end{subequations}
where we have used $\{\{\sub{A}{d},\sub{B}{d}\},\Delta\}=\{\{\sub{A}{d},\Delta\},\sub{B}{d}\}$ which follows from the Jacobi identity since $\{\sub{B}{d},\Delta\}=0$.    {For the more common BAB form of the Wisdom-Holman maps~(\ref{eq:WH:maps}) we thus find from equations~(\ref{eq:WH:Herr:BAB}) and (\ref{eqs:h2:diff}) for the difference in the error Hamiltonian}.
\begin{equation}
	\delta\sub{H}{err}\equiv\sub{H}{err}^{\mathrm{WHI}}-\sub{H}{err}^{\mathrm{WHD}}=
	\frac{h^2}{24} \{\{\sub{A}{d},\Delta\},2\sub{A}{d}+\Delta\}
	+\mathcal{O}(h^4).
\end{equation}
From
\begin{subequations}
\begin{eqnarray}
	\{\{\sub{A}{d},\Delta\},\sub{A}{d}\}
	&=& \frac{1}{M} \sum_{i\neq0}\frac{Gm_0}{Q_i^5}
		\left[(\vec{P}_0\cdot\vec{P}_{i})Q_i^2
			-3(\vec{P}_0\cdot\vec{Q}_i)(\vec{P}_{i}\cdot\vec{Q}_i)\right]
	=	\phm \sum_{i\neq0}\frac{Gm_0m_i}{r_{i0}^5}
		\left[(\sub{\vec{v}}{\!cm}\cdot\tilde{\vec{v}}_{i})r_{i0}^2
			-3(\sub{\vec{v}}{\!cm}\cdot\vec{x}_{i0})
			  (\tilde{\vec{v}}_{i}\cdot\vec{x}_{i0})\right],
	\\
	\{\{\sub{A}{d},\Delta\},\sub{B}{d}\}
	&=&
		\frac{1}{M} \sum_{i\neq0}\frac{Gm_i}{Q_i^5}
		\left[(\vec{P}_0\cdot\vec{P}_{\!\odot})Q_i^2
			-3(\vec{P}_0\cdot\vec{Q}_i)(\vec{P}_{\!\odot}\cdot\vec{Q}_i)\right]
	= {-} \sum_{i\neq0}\frac{Gm_0m_i}{r_{i0}^5}
		\left[(\sub{\vec{v}}{\!cm}\cdot\tilde{\vec{v}}_{0})r_{i0}^2
			-3(\sub{\vec{v}}{\!cm}\cdot\vec{x}_{i0})
			  (\tilde{\vec{v}}_{0}\cdot\vec{x}_{i0})\right],
	\\
	\{\{\sub{A}{d},\Delta\},\Delta\}
	&=&
		\frac{1}{M^2} \sum_{i\neq0}\frac{Gm_0m_i}{Q_i^5}
		\left[P_0^2Q_i^2-3(\vec{P}_0\cdot\vec{Q}_i)^2\right]
	\qquad\quad\;\;\;\,
	=	\phm \sum_{i\neq0}\frac{Gm_0m_i}{r_{i0}^5}
		\left[\sub{v}{\!cm}^2r_{i0}^2
			-3(\sub{\vec{v}}{\!cm}\cdot\vec{x}_{i0})^2\right],
\end{eqnarray}
\end{subequations}
and comparison with equations~(\ref{eqs:Herr:WHD:split}), we find (since $\sub{\vec{v}}{\!cm}=
\vec{v}_i-\tilde{\vec{v}}_i$)
\begin{equation}
\label{eq:herr}
	\delta\sub{H}{err} = \frac{h^2}{24} \left[
		\{\{V_\odot,\Tp\},\Tp\} - \{\{V_\odot,T_1\}T_1\} \right]
	+\mathcal{O}(h^4),
\end{equation}
and the second order term vanishes when $\v{p}_{\mathrm{tot}} = 0$. 

The WHI (and CH) error for the centre of mass obeys
\begin{equation}
	\label{eq:R:err}
	\tdiff{\vec{R}}{t} = \{\vec{R},\delta\sub{H}{err}\}
	= \frac{h^2}{12M} \sum_{i\neq0}\frac{Gm_0m_i}{r_{i0}^5}
		\Big[\vec{v}_{{i}}{r_{i0}^2}-3(\vec{v}_{{i}}\cdot\vec{x}_{i0})\vec{x}_{i0}\Big]
	+\mathcal{O}(h^4)
	= \frac{h^2}{12M} {\left(\pdiff{\vec{F}_\odot}{t}\right)_{\vec{x}_0}}
	+\mathcal{O}(h^4),
\end{equation}
i.e.\ is equal to the $h^2m_0/12M$ times change in acceleration of the Sun {owing to the motion} of the planets {but not the Sun}.

{In the barycentric frame $\{\vec{Q}_{i\neq0},\Delta\} = \vec{P}_0/M$ vanishes and the heliocentric planetary orbits are identical to those obtained with democratic heliocentric coordinates. Hence, the only casualty of using inertial coordinates in the barycentric frame is the centre-of-mass position.} Of course, th{is} error can easily be rectified {(by shifting each particle by $\delta\vec{x} = [\sub{\vec{R}}{ini}-\vec{R}]/M$)}, even after the map has been applied many times. {Then, } using a barycentric frame, this method obtains a map completely equivalent to WHD.

\subsection{The alternative split in inertial coordinates: WHIS}
\label{sec:WHI:alt}
The idea, explored in Section~\ref{sec:WHD:alt}, to split the kinetic energy such that the Kepler problems have gravitating masses $m_0+m_i$ (rather than $m_0$) and the map becomes exact in the single-planet limit, can also be ported to WHI. To this end, we move $\tilde{T}_{\mathrm{c}}$ from eq. \eqref{eq:movei}, in the current coordinates ($\v{P}_i \rightarrow \v{p}_i$), from $\sub{B}{i}$ to $\sub{A}{i}$ and obtain instead of (\ref{eqs:WHI:A+B})
\begin{equation} \label{eqs:WHI:alt:A+B}
	\sub{\tilde{A}}{i} = \sub{A}{i} + \sub{\tilde{T}}{c} = \Tpt + V_\odot = 
		\sum_{i\neq0}\frac{\vec{p}_i^2}{2\mu_i} - \frac{Gm_0m_i}{r_{i 0}}
	\qquad\text{and}\qquad
	\sub{\tilde{B}}{i} = \sub{B}{i} - \sub{\tilde{T}}{c} = \tilde{T}_\odot + \Vp = 
		\frac{1}{2m_0} \left(\vec{p}_0^2 - \sum_{i\neq0}\vec{p}_i^2\right)
		- \sum_{0<i<j}\frac{Gm_im_{\!j}}{r_{i j}},
\end{equation}
and call the new map WHIS.  We have not resolved the non-barycentric map deterioration.  In a barycentric frame, 
\begin{equation}
\v{p}_0^2 - \sum_{i \ne 0} \v{p}_i^2 = 2 \sum_{0< i< j} \v{p}_i \cdot \v{p}_j,
\end{equation}
and $\sub{\tilde{B}}{d} $ = $\sub{\tilde{B}}{i} $.  WHIS still exactly conserves angular and linear momenta because the piece we shifted from $B$ to $A$ has zero Poisson bracket with the momenta.  The kinetic and potential components of $\sub{\tilde{B}}{d}$ are not in involution (in contrast to those of $\sub{B}{i}$ and $\sub{B}{d}$) such that the order of execution of the corresponding maps matters. 

To obtain the WHIS integrators, error Hamiltonians, and nested Poisson brackets, we make the substitutions 
\begin{subequations}
\begin{align}
\tilde{T}_0 &\rightarrow \tilde{T}_\odot,     
& 
\tilde{T}_1 & \rightarrow {\Tpt}, 
&
\tilde{T}_{\mathrm{d}} & \rightarrow \tilde{T}_{\mathrm{c}},
&
\v{Q}_{i} & \rightarrow \v{x}_{i 0},
&
\v{P}_{i} & \rightarrow \v{p}_i,
&
\v{P}_\odot & \rightarrow -\v{p}_0,
&
\tilde{\v{w}}_i & \rightarrow \v{p}_i/\mu_i,
\end{align}
\end{subequations}
in maps \eqref{eq:altsplit}, error Hamiltonians \eqref{eq:splita} and \eqref{eq:splitb}, and nested Poisson brackets \eqref{eqs:Herr:WH:split}. 

We can again parameterise the difference to WHD as $\sub{\tilde{A}}{i}=\sub{\tilde{A}}{d}+\tilde{\Delta}$ and
$\sub{\tilde{B}}{i}=\sub{\tilde{B}}{d}-\tilde{\Delta}$ with
\begin{eqnarray}
\label{eq:test}
	\tilde{\Delta} = \Delta + \sub{\tilde{T}}{c} - \sub{\tilde{T}}{d}
	=& \displaystyle
	    \Delta + \frac{\sum_{i\neq0}m_i^2/m_0}{2M^2} \vec{P}_0^2 +
		\frac{\vec{P}_0}{M}\cdot\sum_{i\neq0}\frac{m_i}{m_0}\vec{P}_i
		\qquad
	&=  \Delta - \frac{\sum_{i\neq0}m_i^2/m_0}{2M^2} \sub{\vec{p}}{tot}^2 +
		\frac{\sub{\vec{p}}{tot}}{M}\cdot\sum_{i\neq0}\frac{m_i}{m_0}\vec{p}_i
	\\
	=&\displaystyle
		\frac{M-m_0+\sum_{i\neq0}m_i^2/m_0}{2M^2}\vec{P}_0^2+
		\frac{\vec{P}_0}{M}\cdot\sum_{i\neq0}\frac{m_i}{\mu_i}\vec{P}_i
	&= \frac{M+m_0-\sum_{i\neq0}m_i^2/m_0}{2M^2}\sub{\vec{p}}{tot}^2 - 		\frac{\sub{\vec{p}}{tot}}{M}\cdot
		\left(\vec{p}_0-\sum_{i\neq0}\frac{m_i}{m_0}\vec{p}_i\right).
\end{eqnarray}
$\tilde{\Delta}$ depends on the momenta $\vec{p}_{i\neq0}$ not just through the combination $\sub{\vec{p}}{tot}$, but explicitly and hence is not in involution with $\Vp$, unlike the situation for $\Delta$. We have
\begin{eqnarray}
	\{V_\odot,\tilde{\Delta}\} &=&
		\frac{\vec{P}_0}{M}\cdot\sum_{i\neq0}\frac{G(m_0+m_i)m_i}{Q_i^3}\vec{Q}_i,
	\\
	\{\Vp,\tilde{\Delta}\} &=&
		\frac{\vec{P}_0}{m_0M}\cdot\sum_{0<i<j}
		\frac{Gm_im_j(m_i-m_j)}{Q_{i\!j}^3}\vec{Q}_{i\!j},
	\\
	\label{eq:Q:tildeD}
	\{\vec{Q}_{i\neq0},\tilde{\Delta}\} &=& \frac{m_i}{\mu_iM}\vec{P}_0,
	\\
	\{\vec{R},\tilde{\Delta}\} &=& 
	{\frac{M-m_0-\sum_im_i^2/m_0}{M}\vec{P}_0}+
		\sum_{i\neq0}\frac{m_i}{\mu_i}\vec{P}_i
	=
	{
		\sub{\vec{p}}{tot}-\vec{p}_0 + \sum_{i\neq0}\frac{m_i}{m_0}
		\left(\vec{p}_i-2\frac{m_i}{M}\sub{\vec{p}}{tot}\right)
	}.
\end{eqnarray}
{Owing to equation~(\ref{eq:Q:tildeD}), the heliocentric trajectories obtained by integrators resulting from this splitting in the barycentric frame} are identical to those obtained from the corresponding split in democratic heliocentric coordinates. The error in the centre of mass is different from that for the scheme discussed before, but again can be trivially corrected.  Hence, WHIS is equivalent to WHDS in the barycentric frame except in the calculation of the center of mass position.
\subsubsection{What is the best velocity frame for WHI, WHIS, and CH?}
\label{sec:vel}
Of all the WHI $\mathcal{O}(h^2)$ terms actually occurring in the error Hamiltonian only
\begin{subequations}
\begin{eqnarray}
	\label{eq:Herr:WHI:V0T0T0}
	\{\{V_\odot,T_\odot\},T_\odot\} &=& \sum_{i\neq0}
		\frac{Gm_0m_i}{r_{i0}^5}\left[
		v_0^2r_{i0}^2-3(\vec{v}_0\cdot\vec{x}_{i0})^2\right], \qquad \text{and}
	\\ \label{eq:Herr:WHI:V0T0T1}
	\{\{V_\odot,T_\odot\},\Tp\} &=& -\sum_{i\neq0}
		\frac{Gm_0m_i}{r_{i0}^5}\left[
		(\vec{v}_0\cdot\vec{v}_i)r_{i0}^2-
		3(\vec{v}_0\cdot\vec{x}_{i0})(\vec{v}_i\cdot\vec{x}_{i0})\right],
\end{eqnarray}
\end{subequations}
depend on the choice of the velocity reference.  We want to calculate the expectation value over long times of the error terms above.  We will need to assume the following relations: 
\begin{subequations}
\begin{align}
	\left\langle \frac{v_0^2}{r_{i 0}^3}  \right \rangle &= \langle {v_0^2}\rangle \left\langle {\frac{1}{r_{i 0}^3}}\right\rangle, \qquad & \text{(assume } \v{x}_{i 0} \text{ and } \v{v}_0 \text{ are uncorrelated)}
	\\ 
	\left\langle \frac{(\v{v}_0 \cdot \v{x}_{i 0})^2}{r_{i 0}^5}  \right\rangle &= \langle {\v{v}_0 \otimes \v{v}_0}\rangle : \left\langle {\frac{\v{x}_{i 0} \otimes \v{x}_{i 0}}{r_{i 0}^5}}\right\rangle = 
	\langle {v_0^2}\rangle\left\langle {\frac{1}{r_{i 0}^3}}\right\rangle,
	\qquad & \text{(assume } \v{x}_{i 0} \text{ and } \v{v}_{0}\text{ components are uncorrelated, respectively)}
	\\
	\left\langle \frac{\v{v}_0 \cdot \v{v}_{i}}{r_{i 0}^3}  \right\rangle &= 
	\langle {\v{v}_0}\rangle \cdot \langle{\v{v}_i} \rangle \left\langle \frac{1}{r_{i 0}^3}\right\rangle,\qquad & \text{(assume planet $i$ uncorrelated from Solar motion)}
	\\
	\left\langle \frac{(\v{v}_0 \cdot \v{x}_{i0})(\v{v}_i \cdot \v{x}_{i0}))}{r_{i 0}^5}  \right\rangle &= 
	\langle {\v{v}_0}\rangle \cdot \langle{\v{v}_i} \rangle \left\langle \frac{1}{r_{i 0}^3}\right\rangle. 
\end{align}
\end{subequations} 
Then, we find,
\begin{subequations}
\begin{eqnarray}
	\langle \{\{V_\odot,T_\odot\},T_\odot\} \rangle &=& 
	-2 G m_0 \langle v_0^2 \rangle \sum_{i \ne 0} m_i \left\langle\frac{1}{r_{i 0}^3}\right\rangle,
	\\ 
	\langle \{\{V_\odot,T_\odot\},\Tp\} \rangle &=& 
	2 G m_0 \langle \v{v}_0 \rangle \cdot \sum_{i \ne 0} m_i  \langle \v{v}_i \rangle \left\langle\frac{1}{r_{i 0}^3}\right\rangle.
\end{eqnarray}
\end{subequations} 
While we have assumed the orbits are not perfectly Keplerian, it is still reasonable to assume  $\langle \v{v}_0 \rangle = \langle \v{v}_i \rangle = 0.$  In this case the first error term \eqref{eq:Herr:WHI:V0T0T0} is minimal in a barycentric frame while the second \eqref{eq:Herr:WHI:V0T0T1} is zero.  So WHI is expected to incur the smallest energy error in a barycentric frame.  For planetary systems including ours, these assumptions may not hold, but our numerical experiments still suggest the barycentric frame is preferable.  

For WHIS, the only nested Poisson brackets depending on the velocity frame are $\{\{V_\odot,\tilde{T}_\odot\},\tilde{T}_\odot\}$ and $\{\{V_\odot,\tilde{T}_\odot\},\sub{\tilde{T}}{pl}\}$:   
\begin{subequations}
\begin{eqnarray}
	\{\{V_\odot,\tilde{T}_\odot\},\tilde{T}_\odot\} &=& \sum_{i\neq0}
		\frac{G_0m_i}{m_ 0r_{i0}^5}\left[
		(m_0 \v{v}_0 + m_i \v{v}_i)^2 r_{i0}^2-3(m_0 \vec{v}_0\cdot\vec{x}_{i0} + m_i \vec{v}_i\cdot\vec{x}_{i0})^2\right],
	\\
	\{\{V_\odot,\tilde{T}_\odot\},\sub{\tilde{T}}{pl}\} &=& -\sum_{i\neq0}
		\frac{Gm_i^2}{\mu_i r_{i0}^5}\left[
		((m_0 \vec{v}_0 + m_i \v{v}_i)\cdot\vec{v}_i)r_{i0}^2-
		3(m_0 \vec{v}_0\cdot\vec{x}_{i0} + m_i \v{v}_i \cdot \vec{x}_{i0})(\vec{v}_i\cdot\vec{x}_{i0})\right].
\end{eqnarray}
\end{subequations}
We find
\begin{subequations}
\begin{eqnarray}
	\langle \{\{V_\odot,\tilde{T}_\odot\},\tilde{T}_\odot\} \rangle &=& 
	-\frac{2 G}{m_0} \sum_{i \ne 0} \left\langle\frac{1}{r_{i 0}^3}\right\rangle \left(m_0^2 \langle v_0^2 \rangle + m_i^2 \langle v_i^2 \rangle \right), 
	\\ 
	\langle \{\{V_\odot,\tilde{T}_\odot\},\sub{\tilde{T}}{pl}\} \rangle &=& 
	2 G \langle v_0^2 \rangle \sum_{i \ne 0} \frac{m_i^3}{\mu_i} \left\langle\frac{1}{r_{i 0}^3}\right\rangle,
\end{eqnarray}
\end{subequations}
using the additional assumption that the components of $\v{v}_i$ are uncorrelated, which is clearly wrong if the orbit is perfectly Keplerian.  But Earth's eccentricity vector, for example, completes a cycle of precession every 112,000 years.  These WHIS error terms are minimal in a barycentric frame, so we expect the WHIS energy error to be minimal in a barycentric frame.  Recall the CH map is identical to the WHI map, so we will find the same conclusions with it.    

\subsection{The methods of Hernandez and Hernandez \& Bertschinger: H16 and HB15}
\label{sec:H16}
We review a new method, explored by \cite{Hernandez2016} that uses Cartesian coordinates, but unlike WHI and WHIS, does not suffer degradation in a non-barycentric frame.  This method has fewer error terms and smaller error than all methods discussed until now and automatically solves $n=1$ correctly.  It correctly solves the center of mass motion, unlike the other Cartesian methods WHI or WHIS.  There is one drawback: it requires using twice as many Kepler solvers as the other methods discussed until now, which means it will be about twice slower when no parallelization of the Kepler problems are performed.  Another drawback is that the Kepler problems are not in involution as was the case of the previous maps.

We have to solve Kepler problems $K_{i j}$, which now have non-zero Poisson brackets for the first time in this paper.  Define functions
\begin{equation} \label{eqs:H16}
	K_{ij} = \frac{\v{p}_i^2}{2 m_i} + \frac{\v{p}_j^2}{2 m_j} - \frac{G m_0 m_i}{r_{i 0}},
	\qquad
	T_{i} =  \frac{\v{p}_i^2}{2 m_i}, 
	\qquad \text{ and} \qquad
	\psi_h^W = \prod_{i=1}^n \Exp{h \hat{K}_{i0}} \Exp{-h(\hat{T}_i + \hat{T}_0 )}.
\end{equation}
The original H16 integrator (variations are found in \cite{DH16}) is 
\begin{equation}
\label{eq:h16}
\phi_h^{\mathrm{H16}} = \Exp{\frac{h}{2} \hat{T}} \Exp{\frac{h}{2}  {\sub{\hat{V}}{pl}}}
\psi_{h/2}^{\dagger W } \psi_{h/2}^{ W } \Exp{\frac{h}{2}  {\sub{\hat{V}}{pl}}} \Exp{\frac{h}{2}  \hat{T}};
\end{equation}
the required operators all commute with the linear and angular momentum operators so the angular and linear momentum are conserved.  The error Hamiltonian is found using eq. (36c) from \cite{DH16}:   
\begin{eqnarray}
	\sub{H}{err,2}^{\mathrm{H16}} &=& \frac{h^2}{48}
	\left[  - 2 \{\{\sub{V}{pl},\sub{T}{pl}\},\sub{T}{pl} \}   +4    \{\{\sub{T}{pl},\sub{V}{pl}\},\sub{V}{pl} \} +8 \{\{\sub{T}{pl},\sub{V}{pl}\},V_\odot \}+ \{\{T_\odot,V_\odot\},V_\odot \}_3  \right],
	\end{eqnarray}
The error is smaller than that for WHI: the $\mathcal O(\epsilon^2)$ error terms $\{\{V_\odot,T_\odot \},\Tp \}$ and $\{\{T_\odot,V_\odot \},V_\odot\}_2$ as well as the $\mathcal O(\epsilon^3)$ term $\{ \{ V_\odot,T_\odot\},T_\odot\}$ have been eliminated which leads to exact solution of the $n = 1$ problem.  $\{\v{R},\tilde{H} \} = \v{p}_{\mathrm{tot}}$ leading to $d\v{R}/dt = 0$ for the continuous time problem described by $\tilde{H}$, so the center of mass motion is correct.  Nested Poisson brackets at all orders depend only on relative, not absolute, velocities, so performance is not degraded in a non-barycentric frame.  

HB15 \citep{HB15} is an even better map as far as error analysis.  It modifies $\psi_h^W$:
\begin{equation} \label{eqs:HB15}
	\psi_h^W = \prod_{{i,j} ~\text{in some order}} \Exp{h \hat{K}_{ij}} \Exp{-h(\hat{T}_i + \hat{T}_j )},
\end{equation}
and one form of the integrator is
\begin{equation}
\phi_h^{\mathrm{HB15}} = \Exp{\frac{h}{2} \hat{T}}
\psi_{h/2}^{\dag W } \psi_{h/2}^{ W }  \Exp{\frac{h}{2}  \hat{T}}:
\end{equation}
a Kepler solver is used to solve the motion between all particle pairs.  Like H16, the HB15 map still conserves angular and linear momentum exactly, solves $n = 1$, and correctly computes the center of mass motion.  HB15 has fewer error terms and smaller error than H16.  We have \citep{DH16}   
\begin{eqnarray}
	\sub{H}{err,2}^{\mathrm{HB15}} &=& \frac{h^2}{48}
	\left[ \{\{\sub{T}{pl},\sub{V}{pl}\},\sub{V}{pl} \}_3 + 2 \{\{\sub{T}{pl},\sub{V}{pl}\},V_\odot \}+ \{\{T_\odot,V_\odot\},V_\odot \}_3  \right],
\end{eqnarray}
so compared to H16, we have eliminated $\{ \{\Vp,\Tp\},\Tp\} \propto \epsilon^2$ and $\{\{\sub{T}{pl},\sub{V}{pl}\},\sub{V}{pl} \}_2 \propto \epsilon^3 $, and reduced the magnitude of the coefficients of $\{\{\sub{T}{pl},\sub{V}{pl}\},\sub{V}{pl} \}$ and $\{\{\sub{T}{pl},\sub{V}{pl}\},V_\odot \}$.  We will see numerically that the fourth order Hamiltonian is also more accurate for HB15 as compared to H16.  HB15's error Hamiltonian does not depend on the velocity frame at any order; in fact, $H_{\mathrm{err},2}$ does not depend on velocities at all.  Unfortunately, HB15 is expensive and requires $n(n+1)/2$ Kepler solvers and is therefore only recommended if its slower speed is not a limitation.  One way to overcome HB15's slower speed is to parallelize the Kepler solver work \citep{DH16}.    
\section{Comparison Tests}
\label{sec:comp}
\begin{figure}
	\includegraphics[width=160mm]{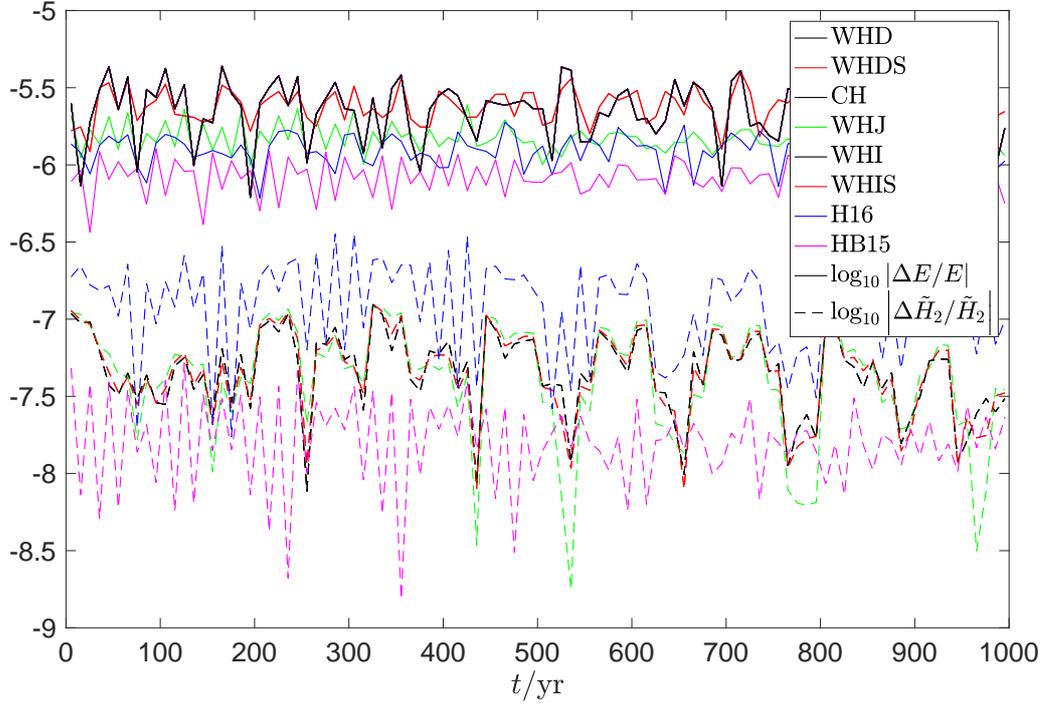}
	\caption{We compare all methods discussed in this paper.  We run a test in a barycentric frame.  We run the outer giant planets problem and calculate the error in energy and second order error Hamiltonian as a function of time.  To reduce error fluctuations, the median error every 10 years is plotted.  As expected, the WHD, CH, and WHI curves are identical, and plotted in the same color, neglecting roundoff error.  WHDS and WHIS are identical and also plotted in the same color.  The lowest errors are given by H16 and HB15. 
		\label{fig:comparebar}
  	}
\end{figure}

A summary of the relative strengths and weaknesses of the integrators in this paper is shown in Table \ref{tab:classify}.  Note that although some methods do not conserve $\v{R}$, this error can be fixed easily as discussed previously.  It appears the method with most advantages is the original Wisdom-Holman method in Jacobi coordinates, WHJ, from these criteria.  However, the method with smallest energy error is H16 and HB15 (however, if we are concerned with exact trajectories during short timescales a high order method such as that of \cite{RS15} is recommended).  
\begin{center}
\begin{table*}
\caption{Summary of methods discussed in this paper.  We write the acronym of the method and the section where the method was introduced.  We indicate whether the method solves the $n=1$ problem and the relative computational expense measured in the number of Kepler solver evaluations per step.  We write whether the method correctly calculates the center of mass position and whether its performance worsens when $\v{p}_{\mathrm{tot}} \ne \v{0}$.  Note that any center of mass error can be corrected as explained in the text.} 
\centering
\begin{tabular}{| l || l | l | l | l | l | l | l | l | l |}
	\hline
	 Method & Section & Solves $n$=1?/Mass entering period equation& Kepler solvers/step (expense) & Conserves CM vector $\v{R}$? & Good performance for any $\v{p}_{\mathrm{tot}}$? \\ [3ex] \hline
	WHD & \ref{sec:WHD}& N/$m_0$ & $n$ &  Y & Y\\ \hline
 	WHDS & \ref{sec:WHD:alt} & Y/$m_0 + m_i$ & $n$ & Y & Y\\ \hline
	CH & \ref{sec:chc}  &N/$m_0$ & $n$ &  N & N\\ \hline
	WHJ & \ref{sec:WHJ} & Y/$m_i + m_0$ or $M_i$ or others &$n$ & Y & Y\\ \hline
	WHI & \ref{sec:WHI} & N/$m_0$ & $n$ & N & N\\ \hline
	WHIS & \ref{sec:WHI:alt} &Y/$m_0 + m_i$ & $n$ &N & N \\ \hline
	H16 & \ref{sec:H16} & Y/$m_0 + m_i$ & $2n$ & Y & Y\\ \hline
	HB15 & \ref{sec:H16} & Y/$m_0 + m_i$ & $(n+1)n$ & Y & Y\\ \hline
	\end{tabular}
\label{tab:classify}
\end{table*}
\end{center}   
In Fig. \ref{fig:comparebar} we show the error in energy and the second order error Hamiltonian, $\tilde{H_2}$ as a function of time for the methods in this paper.  We use the BAB forms of all maps except for H16 and HB15; those integrators are given by equations \eqref{eq:h16} and \eqref{eqs:HB15}, respectively.  A barycentric frame is used and we have integrated the outer giant planets problem with $h = 1$ yr and $t = 1000$ yrs.  The median point every 10 years has been plotted to reduce error oscillations.  

The WHD and CH curves are exactly identical because the coordinates are identical, except for $\v{Q}_0$.  WHI differs from them only at the level of roundoff error.  WHDS and WHIS differ at the level of roundoff as well.  Curiously, H16's $\tilde{H}_2$ error is largest from all methods, but that has no impact on energy error.  We confirm that if one is willing to sacrifice for a method twice as slow (for non-vectorized code), H16 is the best method.  HB15 is even better, but slower.  

\begin{figure}
	\includegraphics[width=160mm]{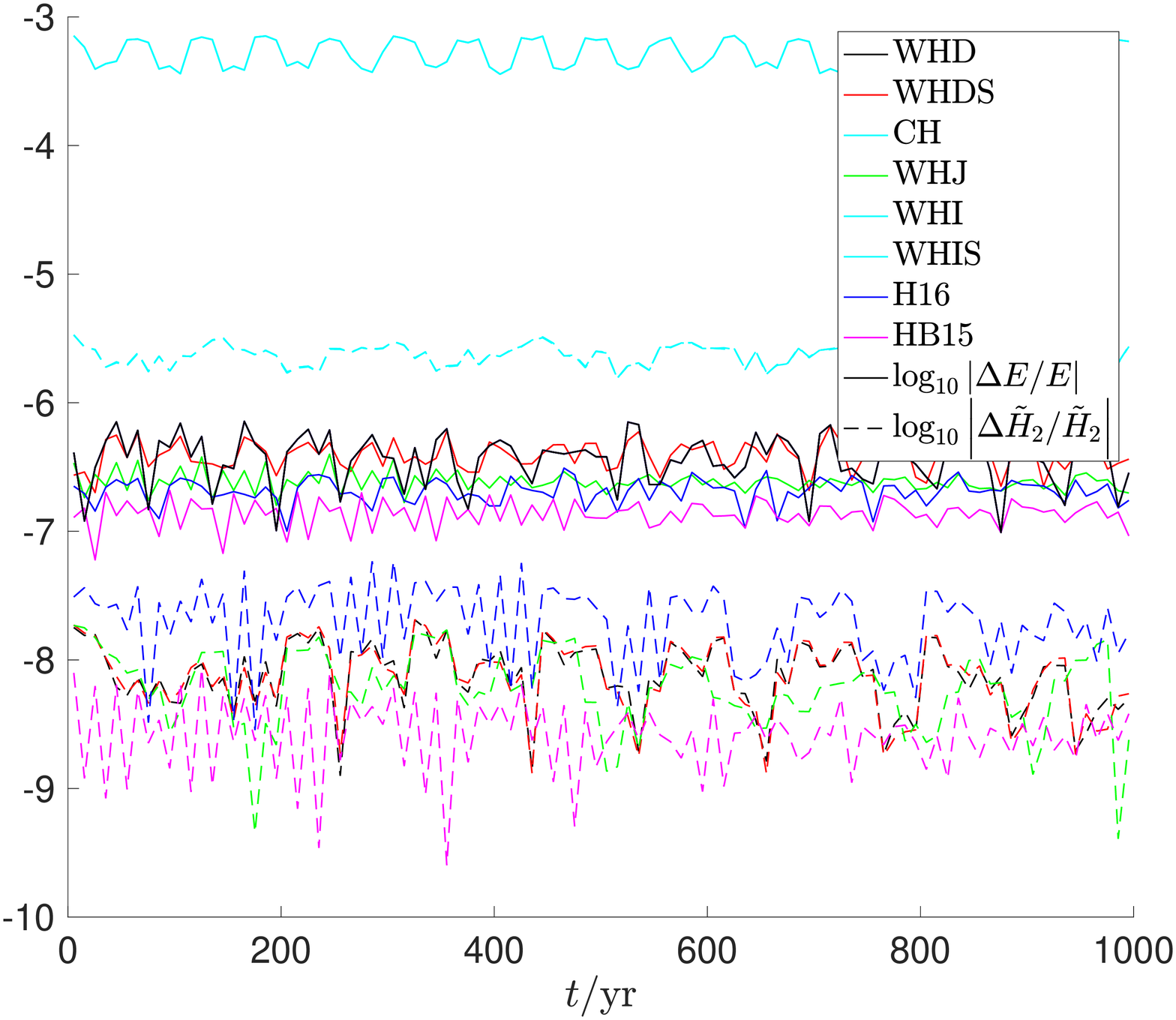}
	\caption{The same test as in Fig. \ref{fig:comparebar} but now using a non-barycentric frame.  The magnitude of the frame's momentum is more than 100 times the magnitude of the RMS momentum vector in a barycentric frame.  The CH, WHI, and WHIS curves have shifted significantly and their performance has deteriorated.  The other curves shift slightly due to roundoff error and different normalizations in the errors.   
	\label{fig:nonbary}
  	}
\end{figure}

We repeat the experiment in a non-barycentric frame to confirm our predictions from Sec. \ref{sec:vel} that some integrators' energy error deteriorate in such a frame.  We use the same frame from Sec. \ref{sec:chc}: the magnitude of the momentum of the frame is more than 100 times greater than the magnitude of the RMS momentum vector.  The result is shown in Fig \ref{fig:nonbary}.   
The only significant shifts in the curves happen for CH, WHI, and WHIS.  The CH and WHI curves differ only by roundoff error.  The WHIS curves differ from the former two by a larger amount, but the amount is indistinguishable in this plot.  The shifts in the other curves are explained by different normalizations in the errors and roundoff behavior.

\begin{figure}
	\includegraphics[width=160mm]{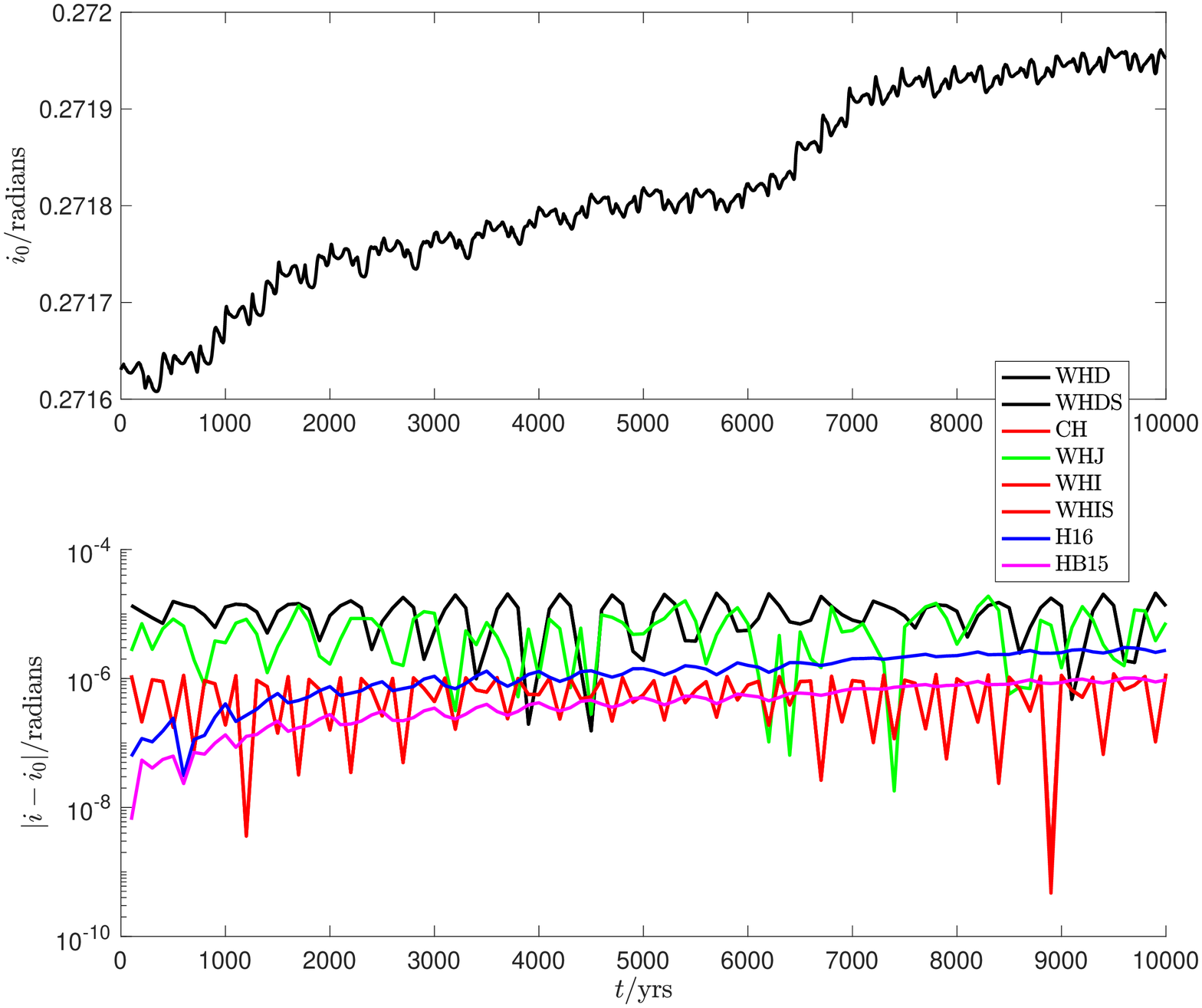}
	\caption{We test how accurately the maps we presented in this paper compute the inclination of Pluto in an $n$-body problem consisting of the Sun, outer gas giants, and Pluto.  The top panel shows the `analytic' inclination $i_0$ as a function of time, calculated by using the H16 map with small step $h = 0.01$ yrs.  The bottom panel shows the errors of the maps used with $h = 1$ yr.  Maps that were previously worst, CH, WHI, and WHIS, are now among the best.  
	\label{fig:pluto}
  	}
\end{figure}

Next we considered the outer giant planets plus Pluto in a barycentric frame.  We calculated the inclination $i$ of Pluto as a function of time for 100,000 yrs.  We ran all maps with $h = 1$ yr.  If $\v{L}_{\mathrm{pluto}}$ is Pluto's angular momentum and $\v{L}_{\mathrm{other}}$ is the angular momentum of the other bodies, we calculate $i$ as
\begin{equation}
i = \arccos \left(\frac{\v{L}_{\mathrm{pluto}} \cdot \v{L}_{\mathrm{other}}}{|\v{L}_{\mathrm{pluto}}| |\v{L}_{\mathrm{other}}|} \right).
\end{equation}
$i$ is usually measured with respect to the ecliptic, which is nearly aligned with the total angular momentum, which itself is nearly due to the gas giant planets; here we are not concerned with the precise $i$ definition to do comparisons for now.  Let $i_0(t)$ be the `analytic' inclination of Pluto.  This is calculated by running H16 with a small time step, $h = 0.01$ yrs which gives a final energy error of $\approx 9.2 \times 10^{-11}$.  We plot $i_0(t)$ in the top panel of Fig. \ref{fig:pluto}.  It is increasing with time on average.    

In the bottom panel we show the error of the inclinations calculated from our maps using $h = 1$ yr.  We plot every 10 points.  There are major changes.  The democratic heliocentric maps (WHD and WHDS) have about an order magnitude worse error than the Cartesian maps WHI and WHIS.  The latter two methods calculate the center of mass position erroneously, but that does not affect calculation of $i$.  The difference between the WHD and WHDS curves is undetectable in this plot and at the level $10^{-9}$; note that the fractional error in gravitating mass for the Pluto Kepler problem is $\approx 7.7 \times 10^{-9}$ when we switch from WHD to WHDS.  The difference between WHI and WHIS is similarly undetectable in this plot.  It appears there is a drift in the HB15 and H16 error curves; actually all curves have a drift but it is most pronounced in these curves, and they do not oscillate around $i-i_0 = 0$.  The relative strengths of the methods have changed compared to Fig. \ref{fig:comparebar}.     

Finally, we test the long term evolution of the orbital elements of Pluto.  The elements of Pluto are of interest, among other reasons, because Pluto's orbit has a large eccentricity, has large inclination, and crosses with Neptune's.  We run the outer giant planets plus Pluto with H16, using $h = 1$ yr and $t = 500$ Myr.  This test was done in \cite{WH91} (hereafter WH91) using WHJ with the same $h$ and $t = 1$ Gyr; we repeat this test but with varying initial conditions.  {As remarked in Section \ref{sec:jac}, Jacobi coordinates are written assuming no planet orbits cross, but Neptune's and Pluto's orbits are crossing- this induces an error.}  We collect sample output every 10,000 yrs as in \cite{Appleetal86}.  The integration lasted under an hour using unvectorized C-code running on a  2.5 GHz Intel Core i7 processor.  

When we use the WH91 initial conditions our results agree with theirs- {the WH91 method is a valid one, as expected}.  But we wish to do a test using currently standard initial conditions in a standard reference frame.  WH91 treated Pluto as a test particle.  We choose masses
\begin{equation}
\begin{aligned}
m_{\odot} &= 1.0, \\
m_{\mathrm{Jupiter}} & = 0.000954786104043, \\
m_{\mathrm{Saturn}} & = 0.000285583733151, \\
m_{\mathrm{Uranus}} & = 0.0000437273164546, \\
m_{\mathrm{Neptune}} & = 0.0000517759138449, \\
m_{\mathrm{Pluto}} & = 6.58086572 \times 10^{-9},
\end{aligned}
\end{equation}
and we verified that if we use the giant planet masses from the Jet Propulsion Laboratory (JPL) DE405 ephemerides our conclusions remain unchanged.  Our rectangular coordinate ephemerides were generated by the JPL HORIZONS system (\texttt{ssd.jpl.nasa.gov/horizons.cgi}) using output date September 5, 1994.  The coordinate system has $\hat{z}$ perpendicular to the plane of the ecliptic and $\hat{x}$ in the direction of the mean equinox of the epoch J2000.0.  This is the conventional coordinate system for calculating Keplerian elements.  We shifted the coordinates into a barycentric frame.  The WH91 initial rectangular coordinate ephemerides are written in \cite{Appleetal86}; their $\hat{z}$ is defined by the total angular momentum vector of the Solar System at some time. 

\begin{figure}
	\includegraphics[width=160mm]{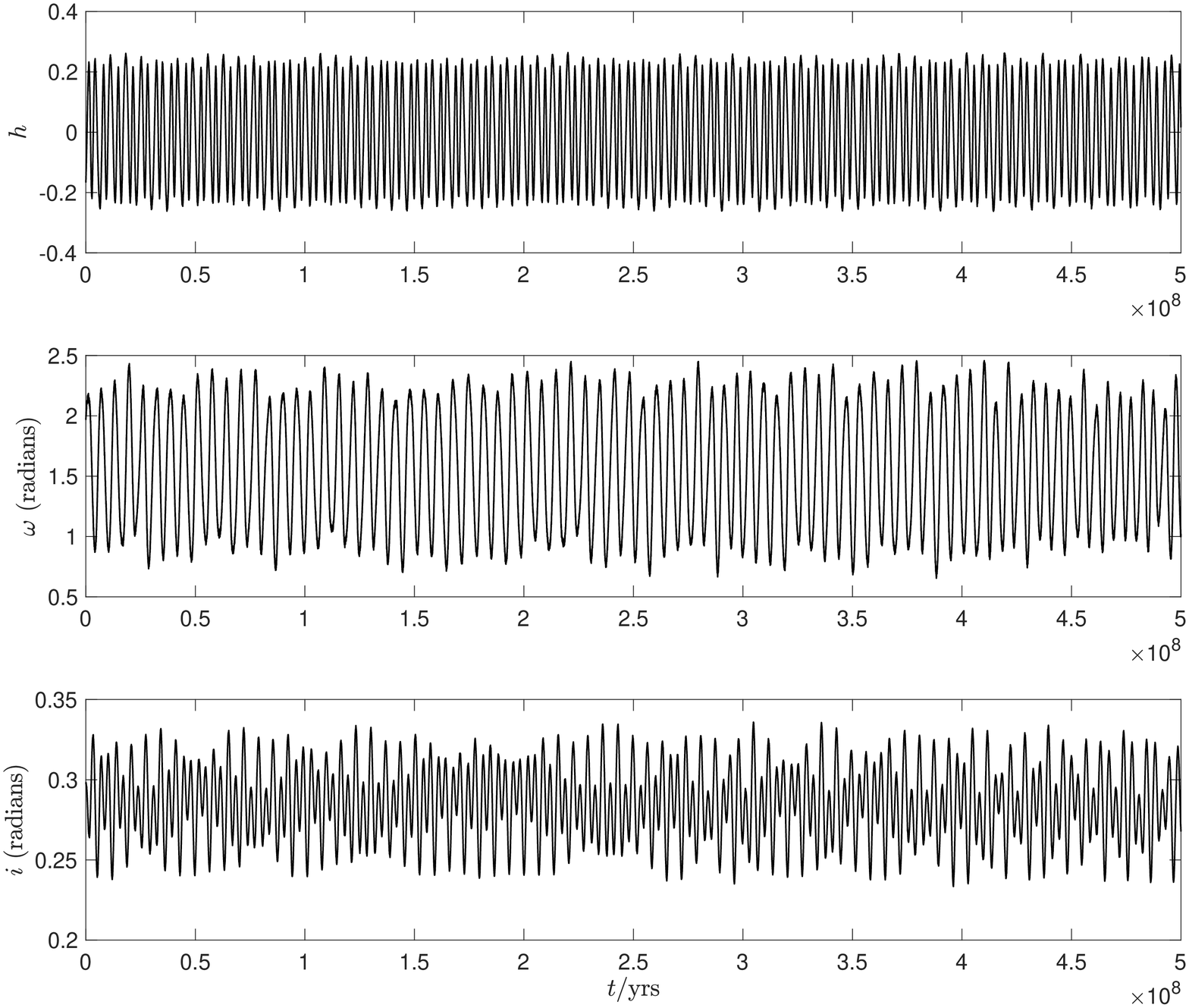}
	\caption{Three of the six Pluto orbital elements tracked over 500 million years.  $h = e \sin(\tilde{\omega})$, $\omega$ is the argument of perihelion, and $i$ the inclination.  We have integrated the Sun, outer gas giants, and Pluto with H16 with step of 1 yr in a repeat experiment of the same test of \protect\cite{WH91} which used WHJ.  Modulations in $h$ are apparently absent in our result and there are other differences.  The differences are due to different initial conditions; for example, our initial conditions use the conventional coordinate frame defined by the ecliptic and equinox.  
	\label{fig:plutolong}
  	}
\end{figure}

Fig. \ref{fig:plutolong} shows three of Pluto's Keplerian elements as a function of time.  We follow the procedure in \cite{dan88} to calculate these elements.  Letting $\tilde{\omega}$ be the longitude of perihelion, the elements are $h = e \sin(\tilde{\omega})$, the argument of perihelion $\omega$, and the inclination $i$.  We have reused the symbol $h$ for this test to match our notation with that of WH91.
The first and third panels can be compared with WH91.  We do not observe the strong 137 million period in $h$ noted in WH91 and the range of $h$ has diminished.  The range of $i$ is larger by about a factor of two.  There is no longer a clear 34 Myr modulation in the $\omega$ behavior.

Using H16 amounts to modifying high frequency terms of the Outer Solar System $N$-body Hamiltonian (WH91).  The frequencies in question are simply the inverse of the step (we revert to usual notation) $h$ ($\approx 32$ nHz in our test).  The averaging principle says that reducing the step size will not affect our conclusions above about Pluto's elements, which we checked by rerunning the test with a step-size three times smaller.  We also repeated the test using HB15 with a step size of 1 yr.   
\section{Conclusions}
\label{sec:conc}
We have studied and done error analyses of the major planetary symplectic integrators in the literature.  These methods use various canonical coordinate systems: Democratic Heliocentric, Canonical Heliocentric, Jacobi, or Cartesian coordinates.  We studied the Wisdom-Holman method in Jacobi \citep{WH91}, Democratic Heliocentric \citep{DuncanLevisonLee1998}, and Cartesian (a new method to our knowledge) coordinates; two methods which correct the single planet limit of the latter two methods; the Canonical Heliocentric map \citep{LaskarRobutel1995}, and two new methods, HB15 and H16, from \cite{HB15} and \cite{Hernandez2016}, respectively.      

We quantified the difference among the maps and explained why some underperform in a non-barycentric frame.  We explained and tested the maps' dependences on $\epsilon$, the ratio of the typical planet mass to the dominant mass {and showed the correct scaling relations, which are not typically discussed}.  We showed that some methods do not correctly calculate the center of mass position and explained why.  We have {restored the full Hamiltonian to form a map which replaces that of} \cite{LaskarRobutel1995}{; this map gives the same solution as a Wisdom-Holman method in inertial Cartesian coordinates that we derive, WHI}.  We compared the errors in energy of the methods when applied to the outer gas giants problem.  Then we included Pluto into the problem and tested how well the maps calculate Pluto's inclination as a function of time.    

The favorable methods from our tests paper are the Wisdom-Holman method in Jacobi coordinates \citep{WH91} and H16 \citep{Hernandez2016}.  The latter uses straightforward Cartesian coordinates and is simpler to implement, while the former usually requires transformations to and from Jacobi coordinates during integration.  We use the latter to perform a new calculation of the evolution of the orbital elements of Pluto over 500 million years. 

We also remark it is easy to generalize the H16 method to more complex planetary systems without a dominant mass without coordinate changes \citep{Hernandez2016}.  It is possible to do such a generalization with WHJ, but the Jacobi coordinates must be modified to hierarchical Jacobi coordinates \citep{SW01}.  In addition, WHJ will not accommodate planetary systems that qualitatively change their nature, such as major bodies that merge, unlike HB15 \citep{HB15}, an extension of H16.  
\section{Acknowledgements}
DMH thanks Edmund Bertschinger for encouragement and numerous helpful discussions, and thanks Jack Wisdom for feedback.  {We thank Dan Tamayo for an interesting and detailed referee report.}
\bibliographystyle{mnras}
\bibliography{refs}

\end{document}